\documentclass[12pt,letterpaper,a4paper]{article}
\usepackage[includeheadfoot,
            marginratio={1:1,2:3},
            width=435pt,
            height=700pt,]{geometry}
\usepackage{amsmath}
\usepackage{amsfonts}
\usepackage{amssymb}
\usepackage{graphicx}
\usepackage{multirow}
\usepackage{empheq}
\usepackage{color}
\usepackage{hyperref}
\usepackage{float}
\restylefloat{table}

\newcommand{\nc}{\newcommand}
\nc{\beq}{\begin{equation}}
\nc{\eeq}{\end{equation}}
\nc{\bea}{\begin{eqnarray}}
\nc{\eea}{\end{eqnarray}}

\def\cO{{\cal O}}
\def\IP{\mathbb{P}}
\def\cO{\mathcal{O}}

\def\cE{\mathcal{E}}

\def\clap#1{\hbox to 0pt{\hss#1\hss}}

\newdimen\csize\csize=1.5ex
\def\young#1{\tiny\vcenter{\hbox{\vrule\vtop{\hrule
  \offinterlineskip\halign{&\vbox
  {\hbox to\csize {\strut\hss##\hss\vrule}\hrule}\cr#1 \crcr}}}}}

\begin{document}

\vspace*{-1.5cm}
\begin{flushright}
  {\small
  MPP-2013-183\\
  }
\end{flushright}

\vspace{1.5cm}
\begin{center}
  {\large \bf On Classifying the Divisor Involutions \\ in 
  Calabi-Yau Threefolds
  }

\end{center}

\vspace{0.75cm}
\begin{center}
Xin Gao$^{\dagger, \,  \ddagger}$\footnote{Email: gaoxin@mppmu.mpg.de} and Pramod Shukla$^\dagger$\footnote{Email: shukla@mppmu.mpg.de}
\end{center}

\vspace{0.1cm}
\begin{center}
\emph{
$^{\dagger}$ Max-Planck-Institut f\"ur Physik (Werner-Heisenberg-Institut), \\
   F\"ohringer Ring 6,  80805 M\"unchen, Germany\\
\vskip 1cm
$^{\ddagger}$ State Key Laboratory of Theoretical Physics, \\
Institute of Theoretical Physics,\\ Chinese Academy of Sciences, P.O.Box 2735, Beijing 100190, China } \\[0.1cm]
\vspace{0.2cm}

 \vspace{0.5cm}
\end{center}


\begin{abstract}
In order to support the odd moduli in  models of (type IIB) string compactification,
we classify  the Calabi-Yau threefolds with $h^{1,1}\le4$
 which exhibit pairs of identical divisors, with different line-bundle charges, mapping to each 
 other under  possible
 divisor exchange involutions.
 For this purpose, the divisors of interest  are identified as completely rigid surface,
 Wilson surface, $K3$ surface and
 some other deformation surfaces. Subsequently, various possible
 exchange involutions are examined under the symmetry of Stanley-Reisner Ideal.
 In addition, we search for the  Calabi-Yau theefolds which contain
 a divisor with several disjoint components.
Under certain reflection involution, such spaces also have nontrivial odd components in (1,1)-cohomology class.
 String compactifications on such Calabi-Yau orientifolds with non-zero $h^{1,1}_-(CY_3/\sigma)$
 could be promising for concrete model building in both particle physics and cosmology.
In the spirit  of using such Calabi-Yau orientifolds in the context of LARGE volume scenario,
we also present some concrete examples of (strong/weak) swiss-cheese type volume form.

\end{abstract}

\clearpage


\tableofcontents
\section{Introduction}
\label{sec:intro}

In order to describe the low energy physics in the real four dimensional world,
string theory has to be compactified on
a six dimensional manifold. To obtain supersymmetry in four dimensions, one requires K\"ahlerity and the
Ricci flatness of such manifolds which result in a Calabi-Yau manifold. For a realistic string model,
especially in order to get a global model,
one of the most challenging requirements is to stabilize the moduli
associated with the Calabi-Yau compactifications.
 In the context of type IIB string compactifications (for a review, see \cite{Grana:2005jc,Lust:2006zg}),
there are two classes of mechanism for moduli stabilization; namely the KKLT \cite{Kachru:2003aw}
and the LARGE volume scenario \cite{Balasubramanian:2005zx}. In both of these mechanisms, complex
structure moduli and axio-dilaton are stabilized at tree level by Gukov-Vafa-Witten(GVW) flux contributions to
the superpotential \cite{Gukov:1999ya} (see related work \cite{Taylor:1999ii,Dasgupta:1999ss}) while the K\"ahler moduli are stabilized with the inclusion
of non-perturbative superpotential corrections coming from $E3$-instanton or gaugino condensation
\cite{Witten:1996bn}. In LARGE volume scenarios, a perturbative $({\alpha^\prime}^3)$ correction
to the K\"ahler potential \cite{Becker:2002nn} is balanced against the exponentially suppressed
non-perturbative superpotential corrections, and subsequently leads to exponentially large stabilized
value for the overall volume of the Calabi-Yau.

Nowadays, people are making great effort to combine global issues like
moduli stabilization and tadpole/anomaly cancellation with the local model building issues such
as getting the  GUT or MSSM-like spectrum.
There are several technical aspects which should be taken care of.
One of the crucial issues is the conflict between chirality in
visible sector and survival of the non-perturbative
superpotential contribution \cite{Blumenhagen:2007sm}.
That is, when one considers a visible sector in a compactification scheme, there are charged instanton zero modes
coming from intersection between E3-brane and D7-brane with world-volume flux,  which will
prevent a class of instantons from participating in moduli stabilization.
Several efforts have been made to avoid such a problem \cite{Collinucci:2008sq,Bobkov:2010rf,Grimm:2011dj,Cicoli:2011qg,
Balasubramanian:2012wd,Cicoli:2012vw,Cicoli:2013mpa}.
One way is, not to support the visible sector on
the divisor which gives rise to the non-perturbative superpotential contribution.
Along these lines, models supporting visible sector with D-branes at singularities have been of interest
\cite{Balasubramanian:2012wd,Cicoli:2012vw,Cicoli:2013mpa}.
This singularity arises from D-flatness condition which forces one or more four-cycles  shrink to
zero size if there are no visible sector singlets which can get a non-vanishing VEV to compensate
the Fayet-Ilopoulos(FI) term in the D-term potential.
In this approach, one needs to embed such singularities in a compact Calabi-Yau threefold
which has to contain non-zero odd components in cohomology class $H^{1,1}_-(CY_3/ \sigma )$
under some holomorphic involution $\sigma$.
Another way to alleviate the chirality issue has been argued in a generalized setup which
includes nontrivial involutively odd two-cycles to support instanton flux \cite{Grimm:2011dj} which again
 requires non-zero $H^{1,1}_-(CY_3 / \sigma )$.
More specifically,  equipped
with the fluxed-insatnton contribution, one includes the involutively odd-moduli ($b^a, c^a$) which
arise from the NS-NS field $B_2$ and R-R field $C_2$ in type IIB orientifolds to correct the E3-brane
superpotential and then remove the extra charged zero modes.
These $b^a$
and $c^a$ moduli appear when there is a nontrivial splitting of cohomology $H^{1,1}(CY_3)$
under holomorphic involution $\sigma$, and  are counted by $h^{1,1}_-(CY_3/\sigma)$.
These  odd moduli will appear in the K\"ahler potential as
well as in the superpotential, and hence can help in moduli stabilization.
In \cite{Gao:2013rra}, the moduli stabilization of these odd moduli is studied in
detail in the context of LARGE volume scenario.
So in either approach, finding Calabi-Yau threefolds with $h^{1,1}_-(CY_3/\sigma) \neq 0$ is
a crucial ingredient, and  is very useful in realizing some particular extensions of  LARGE volume
scenario with the inclusion of odd moduli, which could be promising for both particle physics and cosmology.
This is the main purpose of this paper.

In the context of toric geometry,
the simplest way to have nontrivial (1,1)-cohomology in the involutively odd sector is to exchange
pairs of ``Nontrivial Identical Divisor(NID)"  by requiring  $\sigma: x_i \leftrightarrow x_j$, where $x_i$ is
the homogeneous coordinates and the divisor $D_i\equiv \{x_i=0\}$ is dual to the two-form $\hat D_i$.
The holomorphic involution $\sigma: x_i \leftrightarrow x_j$,
exchanging two homogeneous coordinates of the defining toric Calabi-Yau threefold,
implies a pair of new basis divisors $D_\pm \equiv D_i \pm D_j \in H^{2,2}_{\pm}(CY_3)$  respectively.
The two-form $\hat D_- \in H^{1,1}_-(CY_3)$ dual to the divisor $D_-$ will support
the odd moduli $b^a$ and $c^a, \, a=1,\dots h^{1,1}_-$\footnote{In the following section,
we will drop the `hat' for the dual two-form $\hat D_i$.}.

In such cases, the volume of  orietnifold odd four-cycles $\tau_-$ corresponding to the odd divisor $D_-$
will shrink to zero and not appear in the Calabi-Yau volume form which also implies the splitting
$h^{1,1}_+(CY_3/\sigma) + h^{1,1}_-(CY_3/\sigma)$.
Here, the pairs of NIDs means two divisors with different line-bundle charges (also known as 
Gauged Linear Sigma Model(GLSM) charges)
intersecting with the Calabi-Yau hypersurface to the same topological surface with the same Hodge number.
The requirement of ``Nontrivial Identical''
comes from the fact that the two divisors with the same GLSM charge are not distinguishable due to
the same equivalence relations in the toric data.
As a consequence, exchanging two divisors with the same GLSM charge will not affect the hypersurface polynomial and
so will not contribute to $h^{1,1}_-(CY_3 /\sigma)$.
Some examples with involutions in  elliptic fibered Calabi-Yau threefolds  are studied in \cite{2000math......8011D,Ovrut:2003zj}.  Exchange of del-Pezzo divisors in the Calabi-Yau orientifold compactifications with or without K3 fiberation have been studied in \cite{Blumenhagen:2008zz,Cicoli:2011it}.
Using exchange
involutions, a procedure has been developed in \cite{Collinucci:2009uh} for constructing
F-theory
fourfold uplifts of concrete models of perturbative type IIB (with O3/O7) Calabi-Yau orientifolds.

In the standard approach of moduli stabilization in LARGE volume type IIB  orientifolds, many models
have been constructed with reflection involutions  $\sigma: x_i \rightarrow - x_i$.
Usually such a reflection acts trivially on the homology of the Calabi-Yau space and results
in $h^{1,1}_-(CY_3/\sigma) = 0$, therefore does not support odd moduli in the spectrum.
However, there are exceptions if the
Calabi-Yau space is {\it non-toric} in which reflection involutions can result in $h^{1,1}_-(CY_3/\sigma) \ne 0$.
These {\it non-toric } spaces contain  divisor with disconnected
pieces like $\IP^n  \sqcup \dots \sqcup \IP^n$ or $dP_n  \sqcup \dots \sqcup dP_n$.
We will show that under some reflection, these pieces  exchange to each other and split
to $h^{0,0}_+(D)+h^{0,0}_-(D)$, which will also contribute to the equivariant cohomology $h^{1,1}_-(CY_3/\sigma)$.

Motivated by the possibility of promising utilities of odd moduli in the type IIB extended LARGE volume scenario,
our main goal in this article is to present a systematic  classification of the possible
exchangeable involutions for toric Calabi-Yau threefolds,  and reflection for {\it non-toric}
Calabi-Yau spaces with $h^{1,1}(CY_3) \le 4$. Here, it is important to mention that a complete classification of `all' genuine exchange involutions is highly nontrivial, and  our simplistic approach captures only a subset of such an involutions.
For classification purpose, we will scan through the list of Calabi-Yau
hypersurfaces of toric fourfolds encoded as four-dimensional polytopes in Kreuzer-Skarke list
\cite{Kreuzer:2000xy} for $h^{1,1} \le 4$.
We will proceed with maximally triangulating  all these  reflexive polytopes and subsequently,
get the triple
intersection number along with the K\"ahler cone for each of these
triangulations.
Using these toric data, we will calculate all the Hodge numbers for the coordinate
divisors on the hypersurface to see if they contain pairs of NIDs or
divisors with disjoint pieces.
 For all the spaces (both simplicial and non-simplicial)
containing NIDs, we will check whether there exists divisor exchange involution which is
consistent with the symmetry of the Stanley-Reisner Ideal (SR-Ideal).
Then we will also calculate the volume form to see such splitting $h_+^{1,1}+h_-^{1,1}$ under such involution.
For {\it non-toric} spaces, the volume form contains less
information due to the {\it non-toric} property.  We will directly calculate the
representation of cohomology $H^{1,1}(CY_3)$ to see under which reflection it will split to
odd part and the possibility to contribute as odd modulus.

The article is organized as follows:
In section \ref{sec:divisor}, we present the type of divisors which we consider
in the later classification and their physical background.
In section \ref{sec:classification}, we perform a classification of Calabi Yau spaces with $h^{1,1}(CY_3) \le 4$ for the possible exchange involutions (and for possible reflection involution for the {\it non-toric} spaces) which could result in $h^{1,1}_-(CY_3/\sigma) \neq 0$.
In section \ref{sec:volumeform}, we present the expression of volume forms for all the
classes of exchangeable involutions discussed  in section \ref{sec:classification}.  Some of
the volume forms presented are of  swiss-cheese type.
In section \ref{sec:nontoric}, we discuss the volume form for the {\it non-toric} Calabi-Yau spaces
and calculate the representation of $H^{1,1}(CY/\sigma)$ explicitly.
In the Appendices (\ref{sec:appendix1}-\ref{sec:non-toric}),
We give sketch of computational tools utilized, and tabulate all the
Calabi-Yau threefolds with
explicit exchange (for $h^{1,1}(CY_3) \leq 3$) and reflection (for $h^{1,1}(CY_3) \leq 4$)
involutions which may result in $h^{1,1}_-(CY_3/\sigma) \neq 0$.
The list for all possible exchange involutions and the other relevant topological data for spaces with $h^{1,1}=4$ are
collected in the external file ``Classification of Calabi-Yau Threefolds with Divisor Exchange Involution''.

\section{The Geometry of Divisors}
\label{sec:divisor}

In order to support the odd moduli in a string compactification, we are looking for certain
Calabi-Yau spaces with $h^{1,1}_-(CY_3/\sigma) \neq 0$  which can either exchange pairs of
``Nontrivial Identical Divisor(NID)''  or contain a
divisor with several disjoint components.
The internal geometries for these divisors play an extremely important role in the process of compactification and
subsequent moduli stabilization mechanisms. The most common divisors  which can be investigated for
exchange involution are completely  rigid divisor,
``Wilson'' divisor and deformation divisor.
We will  classify the divisor exchange involution by these divisors in the next section.

\begin{itemize}
 \item Completely rigid divisor

Completely rigid divisor means a divisor with Hodge numbers given as $h^{\bullet}(D)\equiv\{h^{0,0}, h^{0,1},h^{0,2},h^{1,1}\} = \{1,0,0,h^{1,1}\}$
such that  $h^{1,1} \neq 0$.
Depending on whether  $h^{1,1}(D)$ is larger than 9 or not, these divisors are further distinguished as
  del-Pezzo surfaces $\{$ $\IP^2 \equiv dP_0$ , $dP_{n}$, with $n=1, \dots , 8$ $\}$ and  ``rigid but not del-Pezzo'' surfaces
$\{ dP_{n}$, with $n > 9\}$. The del-Pezzo divisors are obtained after blowing up (a set of) generic points in a ${\mathbb P}^2$. Such divisors can be either shrinkable or non-shrinkable depending on their intersections with the other four-cycles. For shrinkable del-Pezzo divisors, one can always find a basis for a given Calabi Yau threefold such that the only non vanishing intersection of the del-Pezzo divisor is the self-intersection\footnote{This is also the reason that sometimes shrinkable del-Pezzo divisors are called as `diagonal del-Pezzo' while non-shrinkable ones are called as `non-diagonal del-Pezzo' \cite{Cicoli:2011it}.}. This property can be observed in the volume form.

On the physical side, having Euler character equal to one is the sufficient condition to have exactly two
fermionic zero modes and therefore a nonzero contribution
to the superpotential \cite{Witten:1996bn}.
The completely rigid divisors satisfy this condition trivially.
In particular,
the shrinkable del-Pezzo surfaces are very important in string phenomenology.
On the particle pheno side, it is argued that such shrinkable divisors are needed to support the GUT 7-branes
in order to decouple from gravity.
On the cosmology side, it is also helpful to realize the swiss-cheese structure in LARGE volume
scenario \cite{Balasubramanian:2005zx,Gray:2012jy}.
 More recently, in the context of  type IIB orientifold with $h^{1,1}_-(CY_3) \neq 0$,
 the identical shrinkable del-Pezzo
 surfaces result in a singularity, which is crucial for global model building on branes
 at del-Pezzo singularity \cite{Balasubramanian:2012wd,Cicoli:2012vw,Cicoli:2013mpa}.

 \item ``Wilson'' divisor

 ``Wilson'' divisor  means a divisor with $h^{1,0}(D) \neq 0$. Here we foucs on the following
``Wilson'' surface of which the Hodge Diamond is $h^{\bullet}(D)=\{1, h^{1,0},0, h^{1,1}\}$ with $h^{1,0}(D)$,
$h^{1,1}(D) \neq 0$.
 In our scanning, we will  specify a particular ``Wilson'' surface as ``Exact-Wilson''
 divisor like $h^{\bullet}(D)=\{1,1,0,h^{1,1}\}$ with $h^{1,1} (D)\neq 0$.
 The physical significance of such divisor comes from the so-called poly-instanton effect
 \footnote{Poly-instanton means the correction of an Euclidean D-brane instanton action by other D-brane instantons
\cite{Blumenhagen:2008ji, Petersson:2010qu,Cicoli:2011ct,Blumenhagen:2012Poly1,Blumenhagen:2012ue, Gao:2013hn}. }.

 In moduli stabilization, as one has to consider a sum over all possible instanton contributions,
 it is a highly nontrivial task to ensure that the considered corrections are the only possible ones.
 In the context of type IIB orientifolds, it has been shown that in the presence
of  an Exact-Wilson divisor with $h^{1,0}_+(D)=1$, one has the right zero mode structure for an Euclidean D3-brane wrapping on it
to generate poly-instanton
 effect in the superpotential \cite{Blumenhagen:2008ji, Blumenhagen:2012Poly1}.
 Such new non-perturbative effects will result in a new class of K\"ahler moduli inflation
 called Poly-instanton inflation
 which treats the ``exact-Wilson'' divisor volume (or together with its axion part)
 as inflaton.

 \item Deformation divisor

 Deformation divisor means a divisor with $h^{2,0}(D) \neq 0$.
It has been proposed that turning on  world-volume
fluxes, it is possible to lift these (extra) deformation zero modes  \cite{Bianchi:2011qh} while
leaving the poly-instanton zero modes encoded in $h^{1,0}_+(D)$ unaffected.
Since these fluxes can rigidify some deformation divisors,
such circumstances facilitate the moduli stabilization process by introducing
more terms for superpotential contributions. Such a mechanism has been utilized
to build an explicit de Sitter in \cite{Louis:2012nb}.
 Here we focus on the following three kinds
 of deformation divisors.

 {\it K3 surface: }
 The Hodge diamond of a K3 divisor is $h^{\bullet}(D)=\{1,0,1,20 \}$. K3 surface present in a
 Calabi-Yau may or may not be fibered. A K3-fibred Calabi-Yau compactification is very helpful
 to obtain an anisotropic shape of the Calabi-Yau. This leads to some LVS models with
effectively two large extra dimensions of micron size \cite{Cicoli:2011it, Cicoli:2011yy}.
The property of spaces which contain both $K3$ and Wilson surface are also studied in
\cite{Blumenhagen:2012Poly1,Lust:2013kt}.

 {\it Special deformation surface: }
 In our scanning, there are two kinds of deformation divisors which appear very frequently.
 One has an extra  $h^{1,1}$ deformation for the K3 surface as $h^{\bullet}(D)=\{1, 0, 1, 21\}$
 which  will be labeled  as $SD1$.
 The other one has a Hodge diamond of  $h^{\bullet}(D)=\{1, 0, 2, 30\}$ which will be labeled as $SD2$.
 Both of these appear to have similar property in the volume form as K3 surface.

\end{itemize}

Now we turn to the {\it non-toric} spaces.
As we described in the introduction,
 these spaces  contain a divisor which consists of some disconnected
pieces like $\IP^n  \sqcup \dots \sqcup \IP^n$ or $dP_n  \sqcup \dots \sqcup dP_n$.
Under some reflection $\sigma: x_i \leftrightarrow -x_i$, these individual components $D'$ and $D''\, \dots$  exchange to
each other, and the combination of these will split
to $h^{0,0}_+(D)+h^{0,0}_-(D)$.
The rigid disjoint pieces  imply that such splitting will also contribute to the equivariant
cohomology $h^{1,1}_-(CY_3/\sigma)$ on the hypersurface.
In these spaces, the number of K\"ahler cone generators and the number of toric equivalence relations 
are always smaller than $h^{1,1}(CY_3)$, and in the context of
Type IIB orientifold one K\"ahler deformation is {\it non-toric}.

\section{The Classification of Calabi-Yau Spaces}
\label{sec:classification}
In this section, we classify the Calabi Yau threefolds for suitable exchange involutions. For this we follow a two-step approach. First, we search for all divisors which have identical hodge-diamond and different line bundle charges. This is what we call ``Nontrivial Identical Divisor (NID)''. In the second step, we combine all possible divisor exchange involutions and examine whether each of these involutions is consistent with the symmetry of SR-Ideal.

The reason for requiring the SR-Ideal to be invariant under divisor exchange involution is the fact that different Calabi-Yau hypersurfaces coming from different triangulations  are encoded in different SR-Ideals. For requiring the
exchange involution not to affect the triangulation, we demand the same to be a symmetry of the SR-Ideal for consistency\footnote{The condition of exchange involution being a symmetry of SR-Ideal might be stronger than the actual need, however we impose this to be on safe side.}.
Since this SR-Ideal symmetry is imposed on the ambient space and the hypersurface is invariant under involution, it will ensure that this symmetry is also preserved at the level of intersection form and
K\"ahler cone condition on the hypersurface.  
More specifically,  let us consider a simple example with $h^{1,1}(CY_3) = 3$, and suppose that the intersection form of the Calabi Yau hypersurface is written in a basis of divisors $\{D_i, D_j, D_k\}$ as $I_3 \equiv I_3 (D_i, D_j, D_k)$ and the exchange involution is  given as $\sigma: x_i \leftrightarrow x_j$. Then, in order to be a genuine involution, $\sigma$ has to ensure that under the new basis $\{D_+, D_-, D_k\}$, where $D_\pm = D_i \pm D_j$, the intersection numbers with odd number of odd-indices ($\kappa_{kk-}, \kappa_{++-}, \kappa_{k+-}, \kappa_{---}$) should vanish for all $k$ which can constitute an orientifold invariant basis. One can show that  the
K\"ahler cone condition will be consistently satisfied after orientifold involution.
It is also important to note that since the divisors are given by the vanishing locus of some polynomials,
 SR-Ideal symmetry also ensures that we can exchange two deformation surfaces in a consistent way.

Let us illustrate this in a concrete example. Consider a Calabi Yau threefold defined by the following toric data
\begin{table}[H]
  \centering
 \begin{tabular}{|c|ccccccc|}
\hline
     & $x_1$  & $x_2$  & $x_3$  & $x_4$  & $x_5$ & $x_6$  & $x_7$       \\
    \hline
    8& 4 & 1 & 1 & 0 & 0 & 0  & 2   \\
    0 & 0 & -1 & -1 & 0 & 0 & 2  & 0   \\
    8 & 4 & 1 & 1 & 1 & 1 & 0  & 0   \\
    \hline
     & &$K3$&$K3$&$K3$&$K3$&$W$& $W$ \\
     \hline
  \end{tabular}
 \end{table}

\noindent
The Stanley-Reisner ideal reads
$$ {\rm SR}=  \{x_2 x_3, x_4 x_5, x_ 1  x_ 6 x_ 7\} \, $$
Computing the Hodge number, one finds that the divisors $D_2, D_3, D_4$ and $D_5$ are
K3, $D_6, D_7$ are Wilson line bundle.
So the possible nontrivial involutions are
$\{\sigma_1: x_2 \leftrightarrow x_4 , x_3 \leftrightarrow x_5\}$,
$\{\sigma_2: x_6 \leftrightarrow x_7\}$,
$\{\sigma_3: x_2 \leftrightarrow x_4 , x_3 \leftrightarrow x_5, x_6 \leftrightarrow x_7\}$.
One can see that all the three involutions are consistent with the symmetry of SR-Ideal.
Expanding the K\"ahler form as $J=r^i[K_i],$ the K\"ahler cone is given simply by $r^i>0$.
 One can show that the K\"ahler cone is generated by the divisors with following GLSM charges:
\bea
\label{divisorclass}
\{\{0,\,0,\,1\},\,\{1,\,-1,\,1\},\,\{2,\,0,\,2\}\}.\nonumber
\eea
We first analyze the involution $\{\sigma_1: x_2 \leftrightarrow x_4 , x_3 \leftrightarrow x_5\}$.
Under the involution $\sigma_1$ we can choose one orientifold invariant basis to expand the K\"ahler form $J= t_i D_i$, which are $\{D_1, D_2, D_4\}$ (the same for $D_3$, $D_5$).
For the K\"ahler parameters $t_i$ the K\"ahler cone condition  translates into
\bea
\label{kaehlerconecondition}
t_4>0, \, t_2 >0, \, t_1 >0.
\eea
Under the orientifold involution $\sigma_1$, we define $D_{\pm} = D_2 \pm D_4$ and
the corresponding K\"ahler parameter $t_{\pm}$.
Since the K\"ahler form is even under involution, i.e. $\sigma^*(J) = J$ in type IIB string
framework and therefore belong to $H^{1,1}_+(CY_3)$,
this implies that $J=t_1 D_1+ t_+ D_+$ and one should identify $t_2=t_4$ to make $t_-=0$.
This identification is  consistent with K\"ahler cone condition eq.(\ref{kaehlerconecondition}).
The intersection form under this basis reads
\bea
I_3 &=& 32 D_1^3+8 D_1^2D_2+8 D_1^2D_4+2D_1D_2D_4 \nonumber\\
&=& 32 D_1^3 +16 D_1^2 D_+ + 4 D_1 D_+^2 - 4 D_1 D_-^2, \nonumber
\eea
from which the intersection numbers $\kappa_ {++-} $ and $\kappa_{---}$ indeed vanish.
All the results are  consistent with the symmetry of the SR-Ideal.

\subsubsection*{Scanning set-up}

In the following scanning, we will specify the Calabi-Yau spaces with $h^{1,1} \leq 4$ which
are  {\it non-toric} spaces or
contain nontrivial divisor exchange involutions.
We  restrict  to the toric ambient four-fold $X_4$  using the Kreuzer-Skarke list \cite{Kreuzer:2000xy}
of reflexive lattice
polytopes and only consider the exchange of NID pairs as
described in section \ref{sec:divisor}.
The Kreuzer-Skarke list contains 36, 244 and 1197 polytopes whose resulting threefolds
are considered to have $h^{1,1}(CY_3)=2$, 3 and 4 respectively. We consider all the maximal triangulations of such
polytopes  which result in 39, 342 and 2587 Calabi-Yau spaces.

On collecting these data, two things should be mentioned at the very outset.
First, in triangulation, we don't  take account the  points in the dual-polytope which are interior to facets.
This will not course  problem since we are just interested in the Hodge number of
the coordinate divisor on the Calabi-Yau hypersurface and these interior coordinate divisors
don't intersect with the Calabi-Yau.
Second, not all the maximal triangulations will correspond to distinguishable Calabi-Yau surface
since there may exist flop-transitions of the ambient spaces which don't affect the hypersurface.
Here, we don't distinguish them since it happens very rarely for small value of $h^{1,1}(CY_3)$.

The calculation is done by means of a scanning tool \cite{Gao:2013} for toric analysis (for a brief introduction see Appendix \ref{sec:appendix1})
which combines some properties of
cohomCalg \cite{Blumenhagen:2010pv, cohomCalg:Implementation}, PALP \cite{Kreuzer:2002uu, Braun:2012vh}
(with SINGULAR \cite{DGPS} extension) and  the toric variety package of SAGE \cite{sage}.
We collect the resolved vertex data in the reflexive dual-polytope
without interior points of facets from PALP. Then  fully triangulate them
in  SAGE and pick out the maximal triangulations.
Using these toric ambient  data we calculate the Hodge number of the coordinate divisor on the
hypersurface by Cohomcalg\cite{Blumenhagen:2010pv, cohomCalg:Implementation}, K\"ahler cone generators from
SAGE \cite{sage}
and the triple intersection number from SINGULAR \cite{DGPS}.
Based on the cohomology data of all the coordinate divisors we get,
we can figure out which spaces contain NID and which one is {\it non-toric}.
 For all the spaces (both simplicial and non-simplicial)
containing NIDs, we will check
whether the exchange involution is allowed by the symmetry of SR-Ideal, and examine the consistency of orientifold invariant K\"ahler cone condition. 

\subsubsection*{Scanning result}
Now, we present the result of scanning for Calabi-Yau threefolds which permit exchange involution and
reflection. Before imposing SR-Ideal symmetry, 
out of the 39, 342 and 2587 Calabi-Yau hypersurfaces,
there are correspondingly 2, 104 and 1419 spaces which exhibit  NIDs. 
From these NIDs appearing in the Calabi-Yau spaces, we can construct all possible divisor exchange involutions
up to four pairs of divisor exchange, i.e. involve at most eight coordinate divisors which is sufficient for
describing toric Calabi-Yau threefolds with $h^{1,1} \leq 4$.
After requiring SR-Ideal symmetry of these involutions, the 
respective number of Calabi-Yau threefolds reduce to 2, 45 and 396
\footnote{For these spaces, we didn't include {\it non-toric} spaces.}(see Table \ref{table:spaces}). 
From these data, we can see the constraint of SR-Ideal symmetry dramatically decreases the
number of suitable spaces which permit
possible exchange involutions.
The number of {\it non-toric} spaces is not affected by these constraints.
There are 0, 1 and 16 spaces for $h^{1,1}=2, 3$ and  4 respectively.

\begin{table}[H]
  \centering
  \begin{minipage}{15cm}
  \begin{tabular}{|c|c||c|c||c|}
  \hline
   {$h^{1,1}$} in & \multirow{2}{*}{$\sharp$ of Polytopes} & {$\sharp$ of resulting $CY_3$ }&
   {$\sharp$ of $CY_3$  } & $\sharp$ of constrained $CY_3$   \\
  Kreuzer-Skarke list  &  & (Max.Tri) & w. NIDs & w.NIDs     \\
    \hline
    \hline
    2 & 36 & 39 & 2\footnote{In fact, these two spaces are identical.} & 2  \\
    \hline
      3 & 244 & 342 & 104 & 45  \\
      \hline
        4 & 1197 & 2587 & 1419 & 396 \\
        \hline
  \end{tabular}
  \caption{Number of $CY_3$ spaces with NIDs  before and after imposing Stanley-Reisner Ideal symmetry. 
  The last column does not contain {\it non-toric} spaces.}
  \label{table:spaces}
  \end{minipage}
  \end{table}

The classification
of exchangeable involutions
after the SR-Ideal symmetry constrains are
classified in Table \ref{identical}.
In this classification, we only consider the NID which corresponds to the surfaces
discussed in section \ref{sec:divisor}, i.e.
completely rigid surface, Wilson surface and some deformation surfaces.
Among these Calabi-Yau spaces, there are some  which
can also exchange several pairs of NIDs  simultaneously.
This  is summarized in Table \ref{identical2}.
The relevant toric data along with the possible exchange involutions for spaces with $h^{1,1}(CY_3)=2, 3$ are summarized
 in Table  \ref{GLSMlist1}-\ref{GLSMlist} of the Appendix \ref{sec:appendix2}.
 For the {\it non-toric} spaces, we find that all the disconnected pieces appearing in the particular divisor are rigid surface.
 The counting of such spaces is
 presented in Table \ref{reflection} and the possible
 reflection under which  $h^{1,1}_-(CY_3/\sigma) \neq 0$ is summarised in
 Table \ref{GLSMlist2} of the Appendix \ref{sec:non-toric}.
  \begin{table}[H]
  \centering
  \begin{tabular}{|c|c||c|c||c||c|c|c|}
  \hline
    \multirow{3}{*}{$h^{1,1}$} &
    \multicolumn{7}{ |c| }{Classification of constrained $CY_3$ with NIDs }  \\
    \cline{2-8}
     & \multirow{2}{*}{Total} &  del-Pezzo  & \multirow{2}{*}{$dP_n$, $n>8$}  &  Wilson  &
    \multirow{2}{*}{K3}  & \multirow{2}{*}{SD1} & \multirow{2}{*}{SD2} \\
    & & $dP_n$, $n \leq 8$ &  & (Exact) &    & & \\
    \hline
    \hline
    2 &  2  & 0 &0 & 0& 0&0  &2 \\
    \hline
      3 &  45 & 7 & 6&4(0) & 35 & 2  & 2 \\
      \hline
        4 &  396 & 151 & 232 & 31(3) & 170 & 26 & 38\\
        \hline
  \end{tabular}
  \caption{Classification of constrained $CY_3$ with NIDs.}
  \label{identical}
\end{table}
\begin{table}[H]
  \centering
  \begin{tabular}{|c|c||c|c|c|c|}
  \hline
    \multirow{3}{*}{$h^{1,1}$} &
    \multicolumn{5}{ |c| }{Classification of constrained $CY_3$ with several pairs of NIDs }  \\
    \cline{2-6}
    & Total  &  del-Pezzo  & del-Pezzo  & K3 \& & del-Pezzo, K3   \\
    & & \& K3 &  \& Wilson(Exact)  & Wilson(Exact)  & \& Wilson(Exact)  \\
    \hline
    \hline
    2 & 2 & 0&0 &0  &0  \\
    \hline
      3 & 45 & 0& 1(0)&3(0) &0  \\
      \hline
        4 & 396 & 2 &21(0) & 6(2) &0 \\
        \hline
  \end{tabular}
  \caption{Classification of constrained $CY_3$ with several pairs of NIDs.}
  \label{identical2}
 \end{table}
 \begin{table}[H]
  \centering
  \begin{tabular}{|c|c|c|c|}
  \hline
  & $h^{1,1}=2$ & $h^{1,1}=3$ & $h^{1,1}=4$\\
        \hline
        \hline
  $\sharp$ of non-toric spaces & 0 & 1 & 16 \\
 \hline
  \end{tabular}
  \caption{Number of {\it non-toric} Calabi-Yau spaces.
 }
  \label{reflection}
 \end{table}

\subsubsection*{The list of  Calabi-Yau spaces with divisor exchange involution}
For $h^{1,1}(CY_3)=4$ we collect the result in the external file ``Classification of Calabi-Yau threefolds with divisor exchange involution''
together with $h^{1,1}(CY_3)=3$  cases.
Here, we provide the relevant assistance for understanding how these data are presented.

For convenience we collect $h^{1,1}(CY_3)=4$ data in
four groups in terms of the number of vertex in the reflexive dual-polytopes, i.e. from 5 to 8.
In each of these groups we first present the toric data of $CY_3$ which permit
nontrivial divisor exchange involutions.
Then, we show the classification of these data as the discussion in Table \ref{identical}
and Table \ref{identical2}.

Now, we give one example to illustrate how to read these data. This example is  in
the first place of the list $h^{1,1}(CY_3)=4$ with 5 dual-vertex polytopes. Naively, it contains 2
basis of inequivalent involutions $\sigma_1: \{x_4 \leftrightarrow x_5\}$, $\sigma_2:\{x_7 \leftrightarrow x_8\}$.
The first  involution  exchanges an identical divisor $dP_7$ and the second involution
exchanges identical divisors of Wilson type. However, after checking the symmetry of SR-Ideal, only
the combination of these two involutions $\sigma:\{x_4 \leftrightarrow x_5,\,x_7 \leftrightarrow x_8\}$
are relevant.
In the list, we present these information as following:
\begin{itemize}
 \item Toric data of relevant $CY_3$
\bea
& & \biggl\{\{x_1, x_2, x_3, x_4, x_5, x_6, x_7,
  x_8\},
  \{ \{x_3, x_4\}, \{x_3, x_5\}, \{x_3, x_6\},\{x_4, x_7\},\nonumber\\
 & & \hskip0.3cm \{x_5, x_8\}, \{x_4, x_5\},\{x_1, x_2, x_6, x_7\},\{x_1, x_2, x_7,x_8\},\{x_1, x_2, x_6, x_8\}\}, \nonumber\\
 & & \hskip0.6cm\{\{0, 2, 1, 2\}, \{0, 2,
    1, 2\}, \{-1, 1, 1, 2\}, \{-1, 1, 0, 1\}, \{0, 0, 0, 1\},\nonumber\\
    & & \hskip0.9cm\{0, 0, 1,
   0\}, \{0, 2, 0, 0\}, \{2, 0, 0, 0\}\} \biggr\} \nonumber
\eea
This is the toric data of this Calabi-Yau space. The first and third brackets are the coordinates
and the corresponding GLSM charge respectively. The second bracket
contains the Stanley-Reisner Ideal obtained by SAGE. The form presented in the list can be directly
used to calculate the cohomology in cohomCalg \cite{Blumenhagen:2010pv, cohomCalg:Implementation}.

\item Consistent involution and corresponding cohomology

The general expression for this part is:
\bea
& &\biggl\{ {\rm Index},\{{\rm Involution \,1},\,{\rm Hodge \,number \,}  \}  ,
\{{\rm Involution \,2},\,{\rm Hodge \,number \,}  \},\dots \biggr\} \nonumber
\eea
In this particular example it reads
\bea
 & &\biggl\{ 1, \, \biggl\{\{ x_4 \rightarrow x_5,  x_5 \rightarrow x_4, x_7 \rightarrow x_8, x_8 \rightarrow x_7 \},\nonumber\\
 & & \hskip0.3cm \{\{\{1,0,0\},\{0,8,0\},\{0,0,1\}\},\{\{1,3,0\},\{3,6,3\},\{0,3,1\}\}\}\biggr\} \biggr\} \nonumber
\eea
The index $1$ means it is the first space in the list.
It contains one type of nontrivial exchange involution $\sigma: \{x_4 \leftrightarrow x_5, x_7 \leftrightarrow x_8\}$ which  is consistent with the symmetry of SR-Ideal and exchanges two $dP_7$ surfaces together with two Wilson surfaces.
One can check that this involution is also consistent with the orientifold invariant K\"ahler cone condition.
$\{\{\{1,0,0\},\{0,8,0\},\{0,0,1\}\},$ $\{\{1,3,0\},\{3,6,3\},\{0,3,1\}\}\}$ represents the Hodge
number of $dP_7$ and Wilson divisor respectively.

 \item Classifications

 In this part, we record the index of the spaces which appear under
 the classification according to Table \ref{identical} and \ref{identical2}.
 As a result, the index $1$  will appear three times in the classification list
 in $h^{1,1}(CY_3)=4$, five vertex in the dual-polytope.

\end{itemize}

\section{Typical Volume Forms for Divisor Exchange Involutions}
\label{sec:volumeform}

In this section, we will present some concrete models under each of the classes
presented in the section \ref{sec:classification}. Our main focus will be in obtaining some simple volume
forms including the strong/weak swiss-cheese types, so that these spaces could be utilized for model building
purpose
in (an extended) LARGE volume scenarios (with the inclusion of odd-axions).
The examples presented here are mostly for $h^{1,1}(CY_3)=3$ as we intend to make all the relevant pieces of
information available in the article itself. Note that, for spaces with $h^{1,1}(CY_3)=4$, one has to refer
the external file.

\subsection{Exchange of completely rigid divisors }
First, we consider the Calabi-Yau spaces which have del-Pezzo coordinate divisors (i.e. divisors $D: x_i = 0$ with $h^{1,0}(D) = 0, h^{2,0}(D)=0$).
In addition, these del-Pezzo divisors should satisfy  the criteria of being ``Nontrivial Identical'' which results in 104 spaces. Imposing the SR-Ideal symmetry, this number
further reduces to 45 spaces among which, 7 spaces contains pairs of del-Pezzos NIDs while 6 spaces have `rigid but not del-Pezzos' exchangeable divisors.
We consider an example to see the volume form before and after the involution in this class of examples.

Let us consider one example which contains both del-Pezzo and rigid but non-del-Pezzo divisors.
This Calabi-Yau threefold is the third example in Table \ref{GLSMlist} and
the toric data is

\begin{table}[H]
  \centering
 \begin{tabular}{|c|ccccccc|}
\hline
     & $x_1$  & $x_2$  & $x_3$  & $x_4$  & $x_5$ & $x_6$  & $x_7$       \\
    \hline
 0 & 1 & -1 & -1 & 0 & 0 & 0  & 1   \\
    6 & -1 & 3 & 2 & 0 & 1 & 1  & 0   \\
    0 & 2 & -2 & -1 & 1 & 0 & 0  & 0   \\   \hline
    & $dP_{14}$& $dP_{14}$ & $dP_{8}$ & $dP_8$ &  & &  \\
    \hline
  \end{tabular}
 \end{table}

\noindent
with Hodge numbers $(h^{2,1}, h^{1,1}) = (45, 3)$ and Euler number $\chi=-84$.
The Stanley-Reisner ideal reads
$$ {\rm SR}=  \{ x_ 1  x_ 4, x_ 2  x_ 3, x_ 3  x_ 4,  x_ 1 x_ 5  x_ 6  x_ 7, x_ 2  x_ 5  x_ 6  x_ 7\} \, $$

Computing the Hodege diamonds, one find that the divisors $D_{1,2}$ are rigid surfaces $dP_{14}$ while $D_{3,4}$ are  del-Pezzo $dP_8$ surface. The possible involution, which keeps the SR-Ideal symmetry is
$\sigma:\{x_1 \leftrightarrow x_2, \& \, x_3 \leftrightarrow x_4 \}$.

 Expanding the K\"ahler form as $J=r^i[K_i],$, the K\"ahler cone is given simply by $r^i>0$.
 One can show that the K\"ahler cone is generated by the divisors with following GLSM charges:
\bea
\label{divisorclass}
\{\{0,\,1,\,0\},\,\{-1,\,3,\,-1\},\,\{-1,\,3,\,0\},\,\{0,1,1\}\}
\eea
Here we see that the number of K\"ahler cone generators is larger than $h^{1,1}(CY_3)$, this shows that the polytope is non-simplicial.

The triple
 intersection form  under the basis of smooth divisors as $\{D_3, D_4, D_5\}$ reads
 \bea
 I_3=D_3^3+D_4^3-D_3^2 D_5-D_4^2 D_5+D_3 D_5^2+D_4 D_5^2+D_5^3.
 \eea
 Writing the K\"ahler form in the above basis of divisors as $J=t_3 D_3+t_4 D_4+t_5 D_5$, the resulting volume
 form in terms of two cycle volumes $t_i$ takes the form
 \bea
 {\cal V}=\frac{1}{3 !}\int_{\cal M} J \wedge J \wedge J = \frac{1}{6}(t_3^2+t_4^3 - 3 t_3^2 t_5- 3 t_4^2 t_5
 +3 t_3 t_5^2 + 3 t_4 t_5^2 + t_5^3)
 \eea
 Then the K\"ahler cone eq.(\ref{divisorclass}) can be expanded under these basis as:
 \bea
K_1 = D_5, \, K_2=D_3 +D_5, \, K_3=D_3+D_4+D_5, \, K_4=D_4+D_5.
 \eea
 For the K\"ahler parameters $t_i$ this translates into
 \bea
t_4>0,\, t_5 >0,\, t_3+t_4-t_5>0
 \eea
 Defining the four-cycle volumes $\tau_i=\frac{1}{2} \int_{D_i} J \wedge J$, we find
 \bea
 \tau_3=\frac{1}{2}(t_3-t_5)^2,\, \tau_4=\frac{1}{2}(t_4-t_5)^2, \,
 \tau_5=\frac{1}{2}(-t_3^2-t_4^2+2t_3t_5+2t_4t_5+t_5^2).
 \eea
Taking into account the K\"ahler cone constraints, the volume can be written in the strong swiss-cheese form
\bea
{\cal V}=\frac{\sqrt{2}}{9}(\sqrt{3}(\tau_3+\tau_4+\tau_5)^{3/2}-3 \tau_3^{3/2}-3\tau_4^{3/2})
\eea
The above volume form shows that the large volume limit is given by $\tau_5 \rightarrow \infty$
while keeping the other four-cycles small. This also indicates that the $dP_8$ divisor are shrinkable to
a point in Calabi-Yau hypersurface.\\

{\it Orientifold Projection:}\\

Under the orientifold involution $\sigma: \{x_1 \leftrightarrow x_2 , \,x_3 \leftrightarrow x_4\}$,
we define $D_{\pm}=D_3 \pm D_4$ and then the intersection form in the new basis $\{D_{\pm},D_5\}$ becomes
\bea
I_3= D_5^3+2 D_5^2 D_+ -2 D_5 D_+^2 + 2 D_+^3-2D_5 D_-^2 +2D_+ D_-^2.
\eea
Under orientifold involution, the K\"ahler form is even, i.e. $\sigma^*(J)=J$ and therefore, it must belong
to $H^{1,1}_+(CY_3)$.
The involution condition implies that $t_3 = t_4$ and then we can expand the K\"ahler form in the basis of
divisors as $J=t_+ D_+ +t_5 D_5$ and write down
the four-cycle volumes $\tau_i$ in terms of these two-cycle volumes $t_i$.
Subsequently, the orientifold invariant volume form become
\bea
{\cal V} & = & \frac{1}{6}(t_5^3+6t_5^2 t_+- 6t_5 t_+^2+ 2t_+^3) \nonumber\\
   & = & \frac{1}{9}(\sqrt{6}(\tau_5+\tau_+)^{3/2} -3 \tau_+^{3/2})
\eea
where
 \bea
 \tau_+ =(t_5-t_+)^2, \, \tau_5=\frac{1}{2}(t_5^2+4t_5 t_+-2t_+^2).
 \eea
This is a perfect example for the simplest generalization of the standard LARGE volume model with the
inclusion of a single involutively odd modulus \cite{Gao:2013rra}.

\subsection{Exchange of Wilson divisors with $h^{1,0}(D) \neq 0$}
In this section, we consider the spaces which have  Wilson
coordinate divisors. In addition, these Wilson divisors should satisfy the criteria of being ``Nontrivial Identical'' and also involution should keep the SR-Ideal invariant.
For $h^{1,1}(CY_3)=3$,
only 4 spaces contain pairs of NIDs with $h^{1,0}(D) \neq 0$.
These are numbered as $\{1, 2, 9, 35\}$ in Table \ref{GLSMlist}.
From the application point of view, we consider example in which the Calabi-Yau volume expression can be
written in a simple form. One such example is the second spaces in the Table \ref{GLSMlist}
for which the toric data is:

\begin{table}[H]
  \centering
 \begin{tabular}{|c|ccccccc|}
\hline
     & $x_1$  & $x_2$  & $x_3$  & $x_4$  & $x_5$ & $x_6$  & $x_7$       \\
    \hline
  0 & 0 & 0 & -1 & -2 & 0 & 0 & 3   \\
  0 & 0 & 0 & -2 & -1 & 0 & 3 & 0   \\
 6  & 2 & 1 & 1 &    1 & 1& 0 & 0   \\   \hline
   & & & $dP_8$ & $dP_8$ &  & $W$ & $W$ \\
    \hline
  \end{tabular}
 \end{table}
\noindent
and has Hodge numbers $(h^{2,1}, h^{1,1}) = (45, 3)$ with Euler number $\chi=-84$.
The Stanley-Reisner ideal reads
$$ {\rm SR}=  \{ x_ 3  x_ 4, x_ 3  x_ 7,  x_ 4  x_ 6, x_1 x_2 x_5 x_6, x_1 x_2 x_5  x_ 7\} \, $$
After looking at the internal structure, one finds that the divisors $D_{3,4}$ are  $dP_8$ surfaces
while the divisors $D_{6,7}$ are `Wilson' divisor. Here, the Wilson divisor W is characterized by
$\{h^{0,0}, h^{1,0}, h^{2,0}, h^{1,1}\} \equiv \{1, 4, 0, 2\}$.
 The possible involution which permutes the two Wilson divisor $W$ and also respects the SR-Ideal symmetry, is defined as;
$\sigma:\{ x_3 \leftrightarrow x_4\, \& \, x_6 \leftrightarrow x_7\}$ .

Expanding the K\"ahler form as $J=r^i[K_i]$, the K\"ahler cone defined by $r^i>0$ is non-simplicial, and it is
generated by the divisors with following GLSM charges:
\bea
\label{divisorclassW}
\{\{0,\,1,\,1\},\,\{-3,\,-3, \,3\},\,\{-2,\,-1,\,2\}, \, \{-1,-2,2\}\}
\eea

Let us consider the basis of divisors $\{D_1, D_6, D_7\}$ which is the only basis compatible with an orientifold
invariant K\"ahler cone.  The intersection form in this basis reads
 \bea
 & & I_3= 8 D_1^3 - 12 D_1 D_6^2 - 24 D_6^3 + 6 D_1 D_6 D_7 \\
 & & \hskip1in + 3 D_6^2 D_7 - 12 D_1 D_7^2 +
 3 D_6 D_7^2 - 24 D_7^3. \nonumber
 \eea
 Writing the K\"ahler form in the above basis of divisors as $J=t_1 D_1 + t_6 D_6 + t_7 D_7$,  the K\"ahler cone
 generators eq.(\ref{divisorclassW}) can be written out as:
 \bea
& K_1 =& \frac{D_1}{2}, \, \, K_2= \frac{3}{2} D_1 - D_6  - D_7, \, \\
& K_3= & D_1 -\frac{1}{3} D_6 - \frac{2}{3} D_7, \, \, K_4= D_1 -\frac{2}{3}D_6 -\frac{1}{3} D_7. \nonumber
 \eea
For the K\"ahler parameters $t_i$ this translates into
 \bea
t_1 > 0, \, 3 t_1 > 2 \, (t_6 + t_7), \, 3 t_1 > t_6 + 2\, t_7, \, 3 t_1 > 2 \, t_6 + t_7.
 \eea
Using the intersection polynomial, the  volume
 form in terms of two cycle volumes $t_i$ takes the form
 \bea
 {\cal V} = \frac{4 t_1^3}{3} - 6 t_1 t_6^2 - 4 t_6^3 + 6 t_1 t_6 t_7 + \frac{3 t_6^2 t_7}{2} -
 6 t_1 t_7^2 + \frac{3 t_6 t_7^2}{2} - 4 t_7^3
 \eea
and subsequently, the corresponding four-cycle volumes can be written as
 \bea
 & & \tau_1=4 t_1^2 - 6 (t_6^2 - t_6 t_7 + t_7^2),\, \tau_6= -\frac{3}{2}\, (2 t_6 - t_7) (4 t_1 + 4 t_6 + t_7), \nonumber\\
 & & \hskip1in
 \tau_7=-\frac{3}{2} \, (t_6 - 2 t_7) (4 t_1 + t_6 + 4 t_7).
 \eea
It is not possible to write a `simple' volume form in terms of 4-cycle volumes expressed above. So, we come
to the expression after considering the exchange involution. \\

{\it Orientifold Projection:}\\

Under the orientifold projection $\sigma:\{ x_3 \leftrightarrow x_4\, \& \, x_6 \leftrightarrow x_7\}$,
we define $D_{\pm}=D_6 \pm D_7$ and then the intersection form in the new basis $\{D_1, \, D_{\pm}\}$ becomes
\bea
I_3= 8 D_1^3 - 12 D_1 D_+^2 - 30 D_+^3 - 36 D_1 D_-^2 - 54 D_+ D_-^2
\eea
Again, the K\"ahler form is expanded in the orientifold invariant basis $J=  t_1 D_1+ t_+D_+ $, and then,
the volume form can be written in terms of orientifold invariant 2-cycle volumes
and subsequently, in terms of the 4-cycle volume as below
\bea
{\cal V}&=& \frac{4 t_1^3}{3} - 6 t_1 t_+^2 - 5 t_+^3 \nonumber\\
&=& \frac{1}{18} \left(6 \, (\tau_1 + \tau_+)^{3/2} + \sqrt{3}\, (3 \tau_1 + 2 \tau_+)^{3/2}\right)
\eea
where
 \bea
 \tau_1 =4 t_1^2 - 6 t_+^2, \, \tau_+= -3 t_+ \, (4 t_1 + 5 t_+)
 \eea
Notice that the aforementioned volume form is not of swiss-cheese type.

In fact, we observe that in all the 4 spaces for $h^{1,1}(CY_3) = 3$ (numbered as $\{1, 2, 9, 35\}$
in Table \ref{GLSMlist}) which have involutions defined by the exchange of `Wilson' divisors, we find that
the orientifold invariant volume form expressions take the following forms,
$\{D_1, D_\pm = D_6 \pm D_7\}$,
\bea
{\cal V}&=& a \, \,  t_1^3 -  b\, \,  t_1 t_+^2 - c \, \,   t_+^3 \,
\eea
for some positive values of the constants $a, b$ and $c$.

\subsection{Exchange of Deformation divisors with $h^{2,0}(D) \neq 0 $}
In this section, we consider the spaces which have deformation
coordinate divisors (i.e. divisors $D: x_i = 0$ with $h^{2,0}(D) \neq 0$). In addition, such deformation
divisor should satisfy the criteria of being ``Nontrivial Identical'' as well as should respect the SR-Ideal symmetry. There are 35 such spaces in
$h^{1,1}(CY_3)=3$ list in Table \ref{GLSMlist}.
\subsubsection*{Swiss-cheese example}
Let us consider the Calabi-Yau space (numbered at 21 in the list)  which has the following toric data:
\begin{table}[H]
  \centering
 \begin{tabular}{|c|ccccccc|}
\hline
     & $x_1$  & $x_2$  & $x_3$  & $x_4$  & $x_5$ & $x_6$  & $x_7$       \\
    \hline
 3 & 1 & 1 & 0 & 0 & 0 & 0  & 1   \\
 3 & 0 & 1 & 0 & 0 & 1 & 1  & 0   \\
 3 & 1 & 0 & 1 & 1 & 0 & 0  & 0   \\   \hline
    &   ${\rm SD2}$ & ${\rm SD2}$& &  &  & & $dP_6$ \\
    \hline
  \end{tabular}
 \end{table}
 \noindent
with Hodge numbers $(h^{2,1}, h^{1,1}) = (72, 3)$ and Euler number $\chi=-138$.
The Stanley-Reisner ideal are
\bea
{\rm SR} &=&  \{ x_ 1  x_ 7,  x_ 2  x_ 7,  x_ 1  x_ 3  x_ 4, x_ 2  x_ 5  x_ 6,  x_ 3  x_ 4
x_ 5  x_ 6\} \, \nonumber
\eea
\noindent
where  ${\rm SD2}$ type divisor has Hodge diamond as
$\{h^{0,0}, h^{1,0}, h^{2,0}, h^{1,1}\} \equiv \{1, 0, 2 ,30\}$. Also, the divisors $D_{3,4}$ and $D_{5,6}$
are also of nontrivial identical type Hodge diamond data given as
$\{h^{0,0}, h^{1,0}, h^{2,0}, h^{1,1}\} \equiv \{1, 0, 1 , 23\}$\footnote{This type of deformation divisor does not appear frequently, so we did not take care of such NIDs in our scan.}.

The exchange of two SD2 divisors compatible with the SR-Ideal symmetry is given by the involution $\sigma: \{x_1 \leftrightarrow x_2 \, \& \, x_3 \leftrightarrow x_5 \,
\& \, x_4 \, \leftrightarrow x_6\}$. Now, expanding the K\"ahler form as $J=r^i[K_i],\, i=1,2,3$, the K\"ahler cone defined via $r^i>0$ is generated
by the divisors with following GLSM charges:
\bea
\label{divisorclassSD1}
\{\{1,\,0,\,1\},\,\{1,\,1, \,0\},\,\{1,\,1,\,1\}\}
\eea

Focusing on the  involution $\sigma$, the intersection form in the basis of smooth divisors $\{D_1, D_2, D_7\}$ can be written as
\bea
\label{eq:SD2a1}
& & I_3=3 D_1^2 D_2 + 3 D_1 D_2^2 + 3 D_7^3
\eea
 Writing the K\"ahler form in the above basis of divisors as $J=t_1 D_1 + t_2 D_2 + t_7 D_7$,  the K\"ahler
 cone generators eq.(\ref{divisorclassSD1}) can be written out as:
 \bea
& K_1 =& D_1, \, \, K_2= D_2, \,  K_3=  D_1 +D_2 + D_7
 \eea
which results in the following simple constraints
 \bea
t_1 + t_7 > 0, \, \, t_2 + t_7 >0, \, \, t_7 <0.
 \eea
Using the intersection polynomial, the  volume
 form in terms of two cycle volumes $t_i$ takes the form
 \bea
 {\cal V} = \frac{3 \, t_1^2 \,  t_2}{2} + \frac{3\, t_1 \, t_2^2}{2} + \frac{t_7^3}{2}
 \eea
and subsequently, the corresponding four-cycle volumes can be written as
 \bea
 & & \tau_1=\frac{3}{2} \, t_2 (2 t_1 + t_2), \, \tau_2= \frac{3}{2} \, t_1 (2 t_2 + t_1), \, \, \tau_7=\frac{3}{2} t_7^2 \,.
 \eea
Utilizing the K\"ahler cone conditions, one can rewrite the volume form in terms of divisor volume
expressions as under,
 \bea
 {\cal V} = \frac{\sqrt{2}}{9} (2 \tau_1 - \tau_2 + \beta)\, \sqrt{-2 \tau_1 + \tau_2 + 2 \beta} - \frac{\sqrt{2}}{3 \, \sqrt{3}} \, \tau_7^{3/2}.
 \eea
 where $\beta = \sqrt{\tau_1^2 - \tau_1 \tau_2 + \tau_2^2}$. Observe that the negative contribution
 is coming from a  del-Pezzo divisor $D_7 = dP_6$. This volume form looks a bit complicated,
 however, after considering the involution $\sigma$, it reduces to a simple and nice form as
 we will see in a moment. \\

{\it Orientifold Projection:}\\

For the orientifold projection $\sigma: \{x_1 \leftrightarrow x_2 \, \& \, x_3 \leftrightarrow x_5 \,
\& \, x_4 \, \leftrightarrow x_6\}$, we consider a basis $\{D_\pm = D_1 \pm D_2, \, D_7\}$
in which the intersection polynomial becomes
\bea
I_3= 3 D_7^3 + 18 D_+^3 - 6 D_+ D_-^2
\eea
from which, one can easily deduce the orientifold invariant volume form  as under
\bea
{\cal V}&=& \frac{t_7^3}{2} + 3 \, t_+^3 = \frac{1}{9} \left( \tau_+^{3/2} - \sqrt{6} \, \, \tau_7^{3/2}\right)
\eea
where $\tau_+ = 9 \, t_+^2, \, \tau_7= \frac{3\,  t_7^2}{2}$.

\subsubsection*{K3-fibration example}
Let us consider an example in which the volume form reflects the fibration structure as well as the
presence of odd moduli. For this, a relevant Calabi-Yau space (numbered at 8 in the list)  is given
by the following toric data:
\begin{table}[H]
  \centering
 \begin{tabular}{|c|ccccccc|}
\hline
     & $x_1$  & $x_2$  & $x_3$  & $x_4$  & $x_5$ & $x_6$  & $x_7$       \\
    \hline
 4 & 2  & 0 & 0 & 0 & 0 & 1  & 1   \\
 4 & 2 &  1 & 0 & 0 & 1 & 0  & 0   \\
 4 & 2  & 0 & 1 & 1 & 0 & 0  & 0   \\   \hline
    &   & $K3$ &$K3$  &  $K3$ & $K3$ & $K3$ & $K3$  \\
    \hline
  \end{tabular}
 \end{table}
 \noindent

with Hodge numbers $(h^{2,1}, h^{1,1}) = (115, 3)$ and Euler number $\chi=-224$.
The Stanley-Reisner ideal are
\bea
{\rm SR} &=&  \{ x_ 3  x_ 4, \,  x_ 6 x_7, \,   x_ 1  x_2   x_ 5 \} \, \nonumber
\eea
\noindent
After checking the SR-Ideal symmetry, the consistent orientifold involution is
$\sigma:\{ x_3 \leftrightarrow x_6 \, \& \, x_4 \leftrightarrow x_7 \}$. Expanding the K\"ahler form as
$J=r^i[K_i],\, i=1,2,3$, the K\"ahler cone defined via $r^i>0$ is generated by the divisors with following
GLSM charges:
\bea
\label{divisorclassSD2}
\{\{1,\,0,\,0\},\,\{0,\,0, \,1\},\,\{1,\,1,\,1\}\}
\eea

Focusing on the involution $\sigma$, the intersection form in the basis of smooth divisors $\{D_2, D_3, D_6\}$
can be written as
\bea
\label{eq:SD2a2}
& & I_3=2 \, D_2 \, D_3 \, D_6
\eea
 Writing the K\"ahler form in the above basis of divisors as $J=t_2 D_2 + t_3 D_3 + t_6 D_6$,  the K\"ahler
 cone generators eq.(\ref{divisorclassSD2}) can be written out as:
 \bea
& K_1 =& D_6, \, \, K_2= D_3, \,  K_3=  D_2 +D_3 +D_6
 \eea
implying the K\"ahler cone conditions $t_6>0, \, t_3 >0, \, t_2 +t_3 +t_6>0$. Using the intersection
polynomial, the  volume
 form in terms of two cycle volumes $t_i$ takes the form
 \bea
 {\cal V} = 2 \, t_2 \, t_3 \, t_6 = \frac{\sqrt{\tau_2} \, \sqrt{\tau_3} \, \sqrt{\tau_6}  }{\sqrt{2}}
 \eea
and subsequently, the corresponding four-cycle volumes can be written as
 \bea
 & & \tau_2= 2 \, t_3 \, t_6 , \, \tau_3= 2 \, t_2 \, t_6, \, \, \tau_6= 2 \, t_2 \, t_3 \,. \nonumber
 \eea

{\it Orientifold Projection:}\\

For the orientifold projection $\sigma:\{ x_3 \leftrightarrow x_6 \, \& \, x_4 \leftrightarrow x_7 \}$,
we consider a basis $\{D_2, \, D_\pm = D_3 \pm D_6\}$ 
in which the intersection polynomial becomes
\bea
I_3= 4 D_2 \, D_+^2 - 4 \, D_2 \, D_-^2
\eea
from which, one can easily deduce the volume form  as under
\bea
{\cal V}&=& 2 \, t_2 \, t_+^2 =\frac{\sqrt{\tau_2} \,\, \, \tau_+}{2\sqrt{2}}
\eea
where $ \tau_2 = 2 \, t_+^2, \, \tau_+= 4 \, t_2 \, t_+$. This volume form is extremely simple, however,
reflects the interesting features which was intended to be, i.e. K3-fibration structure along with
presence of odd moduli in the volume form which will appear after writing out $\tau_+$ in terms of $N=1$ chiral variables. For obvious reasons, a simplest generalization into a `weak'
swiss-cheese type volume form (useful for the LARGE volume models with a single involutively
odd modulus) requires Calabi-Yau spaces with $h^{1,1} = 4$ distributing one modulus appearance for;
fibre, base, swiss-cheese hole and the odd modulus. We do not present any $h^{1,1}= 4$ examples as we intend
to be simplistic and self-contained within the article from the point of view of availability of topological data. Recall
that for $h^{1,1} = 4$ examples, the list is available in a separate external file.

\section{ $H^{1,1}(CY_3/ \sigma)$ Splitting in Non-toric Spaces}
\label{sec:nontoric}
In this section, we present a special class of  examples in which a reflection involution
$\sigma: x_i \rightarrow -x_i$ can result in nontrivial involutively odd sector
with $h^{1,1}_-(CY_3/\sigma) \ne 0$.
 We will also calculate the
representation of cohomology $H^{1,1}(CY_3)$ directly to see under which reflection it will split to
odd part and contribute to odd modulus.

As a concrete example, let us consider the Calabi-Yau
three-fold expressed as a degree
12 hypersurface in ${\mathbb{WCP}}^4[1,1,1,3,6]$
\footnote{This space is numbered at 1 in the Table \ref{GLSMlist2} given in  Appendix \ref{sec:non-toric},
and has also been analyzed in \cite{Hosono:1993qy} .}.
The relevant toric data are given as,

\begin{table}[H]
  \centering
 \begin{tabular}{|c|cccccc|}
\hline
     & $x_1$  & $x_2$  & $x_3$  & $x_4$  & $x_5$ & $x_6$         \\
    \hline
 4 & 2 & 1 & 0 & 0 & 0 & 1     \\
 12 & 6 & 3 & 1 & 1 & 1 & 0     \\  \hline
    & &  &  &  &  &  ${\mathbb P}^2 \sqcup {\mathbb P}^2$\\
    \hline
  \end{tabular}
 \end{table}

\noindent
and has Hodge numbers $(h^{2,1}, h^{1,1}) = (165, 3)$ with Euler number $\chi=-324$. Here, since $h^{1,1} =3$
exceeds the number of toric equivalence relations, one K\"ahler deformation is non-toric.
The Stanley-Reisner ideal reads
$$ {\rm SR}=  \{ x_ 1 x_2 x_ 6, x_3 x_ 4  x_ 5\} \, $$
This Calabi-Yau has one  divisor $D_6$ which contains two disjoint pieces $\IP^2 \sqcup \IP^2 $.
The K\"ahler cone condition is just given by $t_2>0$ and $t_3>0$.
The intersection form written in the basis of smooth divisors $\{D_2,  D_6\}$ is,
\bea
\label{Int-non-toric}
& & I_3=18 D_2^3 + 18 D_6^3
\eea
which results in the following volume form,
\bea
\label{eq:volumeform}
& & \hskip-0.8cm {\cal V}\equiv \frac{1}{3 !} \int_{\cal M} J \wedge J\wedge J  = 3 \, t_2^3 + 3
\, t_6^2 = \frac{1}{9} (\tau_2^{3/2} - \tau_6^{3/2})\eea
where $$ \tau_2 = 9 t_2^2,\,  \tau_6 = 9 t_6^2 . \,$$

\subsubsection*{Explicit computation of  $h^{1,1}_-(CY_3/\sigma)$}

For {\it non-toric} space the volume form contains less information,
 it is hard to say  under
which reflection $h^{1,1}(CY_3)$ will split to odd part.
Now, we will calculate the representation for the $H^{1,1}(CY_3)$ to see  such splitting explicitly.
It can be either  written down by anaylzing the Koszul sequence of bundle-valued Cohomology
\cite{cohomCalg:Implementation}
or by the representation of $H^{1,1}(CY_3)$ in terms of homogeneous coordinates \cite{Gao:2012} for
some simple examples.  Now, we use the first method to illustrate the procedure.

 Since $H^{1,1}(CY_3)=H^{1}(CY_3; T^*_{CY_3}) $, by using the corresponding Koszul sequence for
 the cotangent bundle (see \cite{Blumenhagen:2010pv} and references there), the exact sequence becomes\\
 \begin{table}[H]
 \centering
\begin{tabular}{|c|c|c|}
 \hline
   \multirow{2}{*}{$\cE^*_S$} &  $\cO_S(-2,-6) \oplus \cO_S(-1,-3) $ &  \multirow{2}{*}{$\cO_S^{\oplus 2}$} \\
  &$ \oplus \cO_S(0,-1)^{\oplus 3} \oplus \cO_S(-1,0)$  & \\
 \hline
 $0$ & $0$ & $2$ \\
 \hline
 $3$ & $1$ & $0$ \\
 \hline
 $0$ & $0$ & $0$ \\
 \hline
 $59$ & $61$ & $2$ \\
 \hline
 $0$ & $0$ & $ 0$ \\
 \hline
\end{tabular}
$\Longrightarrow$
\begin{tabular}{|c|c|c|}
 \hline
   \multirow{2}{*}{$\cO_S(-4,-12)$} &   \multirow{2}{*}{$\cE^*_S$} &  \multirow{2}{*}{$T^*_S$} \\
  & & \\
 \hline
 $0$ & $0$ & $0$ \\
 \hline
 $0$ & ${\bf 3}$ & ${\bf 3}$ \\
 \hline
 $0$ & $0$ & $165$ \\
 \hline
 $224$ & $59$ & $0$ \\
 \hline
 $0$ & $0$ & $ 0$ \\
 \hline
\end{tabular}\\
\end{table}
More precisely, the three (3 = 2+1) appearing in $H^{1}(CY_3; T^*_{CY_3})$ comes from $H^{1}(\cE^*_S)$, in which `2' comes
from $H^{0}(\cO_S)$
whose  representation is always a constant
while `1' comes from  $H^1(\cO_S(-1,0))$. Again, using the exact sequence in cohomology for the linebundle $\cO_S(-1,0)$,
we find  the `1' contribution to $H^1(\cO_S(-1,0))$ coming from  $H^2(\cO_X(-5,-12))$ on the
ambient space $X$.
Assuming the restriction map and embedding map in the sequence to be  invariant under involution,
 the parity of $H^{1}(CY_3; T^*_{CY_3})$  depends on the parity of $H^2(\cO_X(-5,-12))$
under reflection. The polynomial representation of $H^2(\cO_X(-5,-12))$ can easily be calculated to
be $\frac{1}{x_1 x_2^2 x_6}$.
So, only under reflection involution $\sigma: x_1 \leftrightarrow -x_1$ this polynomial changes signs and then,
$h^{1,1}(CY_3/\sigma)$  split to $h^{1,1}_+=2$ and $h^{1,1}_-=1$.
One can also show explicitly that the cohomology of linebundle  also splits into $h^{0,0}_+(D_6)=h^{0,0}_-(D_6)=1$
 under such reflection.

In this example, we find that $D_6$ divisor has two disjoint ${\mathbb P}^2$s which are internally exchanged within $D_6$ under the reflection involution
$\sigma : x_1 \rightarrow - x_1$. This is related to the fact that one K\"ahler deformation is non-toric
and there are only two generators of the K\"ahler cone which is smaller than the
number of K\"ahler moduli $h^{1,1}(CY_3)=3$.
This implies the K\"ahler form $J$ (which has
to be even under the involution $\sigma$) can be written in terms of two divisor volumes. Such
a case has been also observed in \cite{Blumenhagen:2012Poly1} while investigating the zero-mode
structure for generating poly-instanton corrections.

\section{Conclusions}
\label{sec:conclusion}
In this paper, we presented a classification of  the toric Calabi-Yau threefolds with $h^{1,1}\le4$ relevant for nontrivial involutively odd sector of (1,1)-cohomology class. For this purpose, we studied two types of involutions;
the first one permutes ``Nontrivial Identical Divisor(NID)'' 
under exchange of coordinates $\sigma : x_i \leftrightarrow x_j$ while the other one is a
reflection $\sigma: x_i \leftrightarrow -x_i$ on {\it non-toric} space.
Both of these types of involutions can result in a non-zero $h^{1,1}_-(CY_3/\sigma)$
for the Calabi-Yau threefolds.  In our scanning,  we search for the spaces with NIDs and find that imposing SR-Ideal symmetry 
on the divisor exchange involutions dramatically decreases number of suitable spaces.
As a result, we get the list of toric data for the Calabi-Yau threefold with $h^{1,1} \leq 4$ which contains
non-zero $h^{1,1}_-(CY_3/\sigma)$ when we restrict to {\it non-toric} spaces and exchange involution for
completely rigid divisors, Wilson divisors and  three kinds of deformation divisors.
Such explicit constructions of Calabi Yau orientifolds provide a suitable background to extend the type IIB setups
(such as KKLT or LARGE volume scenarios) with the inclusion of odd-moduli in the spectrum, and
can be useful for concrete as well as promising model building in particle phenomenology and string cosmology.
It is straightforward to generalize  our result to include more spaces as Calabi-Yau threefolds with $h^{1,1} \geq 5$ as well as Calabi-Yau fourfolds.
It would also be interesting to generalize our method to study the spaces beyond the Kreuzer-Skarke list, like higer dimensional polytopes and CICY manifolds (for the list, see \cite{Candelas:1987kf,Gray:2013mja}).

\section*{Acknowledgement}
We are grateful to Ralph Blumenhagen for several enlightening discussions, clarifications and suggestions.
We are very thankful to Yang-Hui He for helpful discussions and comments on the manuscript.  
We also thank Ross Altman, Volker Braun, Andres Collinucci, Christoph Mayrhofer, Brent Nelson, 
Xu Zhang for helpful discussions. 
XG is supported by the MPG-CAS Joint Doctoral Promotion Programme and PS is supported
by a postdoctoral research fellowship from the Alexander von Humboldt Foundation.

\newpage
\appendix
\section{Computation tools}
\label{sec:appendix1}
There are several packages which can analyze some properties of toric varieties,
but sometimes it is not convenient  for each of them to do a complete calculation.
For example, sometimes PALP fails to triangulate the polytope by itself.
This happens whenever the dimension of the polytope is different from four,
or when the polytope contains points which are interior to facets.
The later one may not cause problem if we just consider the HodgeDiamond of
the coordinate divisor on a Calabi-Yau hypersurface since these interior points divisor
don't intersect with the Calabi-Yau hypersurfaces, so these correspond to null HodgeDiamond.
However, if we want to keep control over the lattice points involved in the triangulation,
especially when we analyze general non-Calabi-Yau hypersurface,
it is important to triangulate these points fully. In fact, this is exactly what SAGE can do.
However, it is very hard for SAGE itself to do some scanning stuff. For getting the
intersection number and intersection form SINGULAR is much
more efficient. On the scanning issue, some previous program can only
contain one of many maximal triangulations and more importantly,
it is a black-box which is hard to change and extend. So, it is worthwhile to combine these package
in a systematic way which can  share some advantages of them and
avoid complicate data transfer across several packages.

In this compact scanning program \cite{Gao:2013}, we try to combine some properties of SAGE, PALP(SINGULAR) into a single
Mathematica file based on python language.
There are two similar scanning procedures
depending on different forms of input:
one is scanning the GLSM charges as input and the other one is scanning the data from Kreuzer-Skarke list.
For the later one, we collect the vertex data in the dual-polytope after blow-up from PALP,
then put these data as input to SAGE for fully triangulations.
Here these vertex data can get collected in
two different ways. One is to include the  points which are interior to facets
 and the other one is ignoring them as PALP did itself.  
 After triangulations,
 we can go further to get the K\"ahler cone generators in terms of line-bundle charges 
and meanwhile utilizing these data back to PALP(SINGULAR) to get the
triple intersection number and intersection form.
Moreover, 
this analyze can also be performed beyond the Kreuzer-Skarke list 
like higer dimensional polytopes and CICY manifolds (for the list, see \cite{Candelas:1987kf,Gray:2013mja}).

\section{List of $CY_3$ for $h^{1,1}_-\neq 0$ Under Divisor Exchange Involutions}
\label{sec:appendix2}
In this section, we provide the toric data relevant for those Calabi-Yau spaces in which
nontrivial identical `coordinate' divisors have been found in the scan.  Note, that we have investigated only coordinate divisors for ``Nontrivial Identical''
criteria and hence, it is essential for us to provide the explicit GLSM
charges along with the SR-Ideal for each space. It is understood that
our scan does not capture all divisors which could be ``Nontrivial Identical'' ;
e.g. those which may come from a combination of GLSM charges of various coordinate divisors.

Here we present the list of results for $h^{1,1}(CY_3)=2$ and $3$. For  $h^{1,1}=4$ please see
the external Mathematica file ``Classification of Calabi-Yau Threefolds with Divisor Exchange Involution''.

\subsection*{$h^{1,1}(CY_3)=2$}
In the scan of 39 CalabiYaus with $h^{1,1} =2$, we find that only one has ``Nontrivial Identical'' coordinate divisors. The toric data for this Calabi-Yau is given in Table \ref{GLSMlist1}
along with the possible exchange involution.
\begin{table}[H]
  \centering
  \begin{tabular}{|c||c|}
  \hline
   No. & Toric data   \\
    \hline \hline
& ${\rm GLSM}: [(1,0), (1, 0), (1, 0), (0, 1), (0, 1), (0,1)]$\\
  1  &  $ {\rm SR}: \{x_1 x_2 x_3, x_4 x_5 x_6 \}$ \, and $x_i \equiv SD2$ \, \, $\forall i$.\\
 & $\sigma: \{ x_1 \leftrightarrow x_4,  x_2 \leftrightarrow x_5,  x_3 \leftrightarrow x_6 \}$ \\
 \hline
 \end{tabular}
\caption{List of $CY_3$ spaces with $h^{1,1} = 2$ for the possibility of $h^{1,1}_-(CY_3/\sigma) \neq 0$
under divisor exchange involutions.}
\label{GLSMlist1}
 \end{table}

\subsection*{$h^{1,1}(CY_3)=3$}
In the collection of the Calabi-Yau threefolds with $h^{1,1}(CY_3) =3$ below, we have seven charge
vectors in GLSM corresponding to each coordinate $x_i$ for $ i \in \{1,2,..,7\}$ along with the SR-Ideal
\footnote{After considering all possible maximal triangulations, there is a possibility that some spaces coming from different polytopes might be repeated. For example, we can see this situation explicitly in $h^{11}=3$ where spaces numbered as 5-8 are the same as those numbered as 39-42. However, in order to use the labeling of spaces consistent with the external file, we count all such spaces.}
\begin{table}[H]
  \centering
  \begin{tabular}{|c||c|}
  \hline
   No. & Toric data   \\
    \hline \hline
& ${\rm GLSM}: [(4, 0, 4), (1, -1, 1), (1, -1, 1), (0,
    0, 1), (0, 0, 1), (0, 2, 0), (2, 0, 0)]$\\
\cline{2-2}
 \multirow{3} {*} {1} &  $ {\rm SR}: \{x_2 x_3, x_4 x_5, x_1 x_6 x_7\}$\\
  & $ \sigma_1: \{x_2 \leftrightarrow x_4\, \&  \, x_3 \leftrightarrow x_5 \equiv K3 \} $; \, $ \sigma_2: x_6 \leftrightarrow x_7 \equiv \{1,3,0,2\}$; $\sigma_3 = \sigma_1\cup \sigma_2$  \\
& K\"ahler Cone generators (KC): \, $\{(0,0,1),(1,-1,1),(2,0,2)\}$ \\
  \hline
& ${\rm GLSM}: [ (0, 0, 2), (0, 0, 1), (-1, -2, 1), (-2, -1, 1), (0, 0, 1), (0, 3, 0), (3, 0, 0) ]$\\
\cline{2-2}
  \multirow{3} {*} {2} &  $ {\rm SR}: \{ x_3  x_4, x_3  x_7, x_4  x_6, x_1  x_2  x_5  x_6, x_1  x_2  x_5  x_7\}$\\
  & $ \sigma: \{x_3 \leftrightarrow x_4 \equiv dP_8 \, \& \,  x_6 \leftrightarrow x_7 \equiv \{1,4,0,2\} \}$ \\
& KC: \, $\{(0,0,1),(-3,-3,3),(-2,-1,2),(-1,-2,2)\}$ \\
  \hline
& ${\rm GLSM}: [ (1, -1, 2), (-1, 3, -2), (-1, 2, -1), (0, 0, 1), (0, 1, 0), (0, 1, 0), (1, 0, 0)]$\\
\cline{2-2}
\multirow{3} {*} {3} &  $ {\rm SR}: \{ x_1  x_4, x_2  x_3, x_3  x_4, x_1  x_5  x_6  x_7, x_2  x_5  x_6  x_7 \}$\\
& $ \sigma: \{x_1 \leftrightarrow x_2 \equiv dP_{14}\, \& \, x_3 \leftrightarrow x_4 \equiv dP_8 \}$ \\
& KC: $\{(0,1,0),(-1,3,-1),(-1,3,0),(0,1,1)\}$\\
  \cline{2-2}
 \hline
 \end{tabular}
 \end{table}

\begin{table}[H]
  \centering
  \begin{tabular}{|c||c|}
  \hline
   No. & Toric data   \\
    \hline \hline
& ${\rm GLSM}: [(-2, 2, 1), (0, 0, 1), (1, 1, 0), (0, 1, 0), (0, 1, 0), (1, 0, 0), (1, 0, 0)]$\\
\cline{2-2}
\multirow{3} {*}  {4} &  $ {\rm SR}: \{ x_ 1  x_ 2, x_ 1  x_ 4  x_ 5, x_ 3  x_ 4  x_ 5, x_ 3  x_ 6  x_ 7, x_ 2  x_ 6  x_ 7\}$ \\
& $ \sigma: \{x_1 \leftrightarrow x_2 \equiv dP_1 \, \& \, x_4 \leftrightarrow x_6 \, \&  \, x_5 \leftrightarrow x_7\equiv \{1,0,1,21\}\} $ \\
& KC: \, $\{(1,1,0),(2,2,1),(0,2,1),(0,4,1)\}$\\
\hline
  & ${\rm GLSM}: [(2, 4, 2), (1, 2, 0), (0, 0, 1), (0, 0, 1), (0, 1, 0), (0, 1, 0), (1, 0, 0)] $\\
\cline{2-2}
5 &  $ {\rm SR}: \{ x_ 3  x_ 4, x_ 5  x_ 6, x_ 1  x_ 2  x_ 7 \}$\\
(39)& $ \sigma_1: \{x_3 \leftrightarrow x_5 \, \& \, x_4 \leftrightarrow x_6 \equiv K3 \}$ \, \, {\rm and }\, \, $x_7 \equiv W $ \\
& KC: \, $\{(0,1,0),(0,0,1),(1,2,1)\}$\\
 \hline
   & ${\rm GLSM}:[(2, 2, 2), (0, 1, 0), (0, 0, 1), (0, 0, 1), (0, 1, 0), (1, 0, 0), (1, 0, 0)]$ \\
\cline{2-2}
 &  $ {\rm SR}: \{ x_ 2  x_ 5, x_ 3  x_ 4, x_ 1  x_ 6  x_ 7 \}$\\
   & $ \sigma: \{x_2 \leftrightarrow x_3\, \& \, x_4 \leftrightarrow x_5 \}$, \, \, \, \, \, $x_i \equiv K3 \, \, \forall i = \{2,..., 7\}$\\
& KC : \, $\{(0, 0, 1), (0, 1, 0), (1, 1, 1)\}$ \\
     \cline{2-2}
6-8  &  $ {\rm SR}: \{ x_ 2  x_ 5, x_ 6  x_ 7, x_ 1  x_ 3  x_ 4 \}$\\
(40-42)& $ \sigma: \{x_2 \leftrightarrow x_6\, \& \,  \, x_5 \leftrightarrow x_7\} $, \, \, \, \, \, $x_i \equiv K3 \, \, \forall i = \{2,..., 7\}$\\
& KC: \, $\{(1, 0, 0), (0, 1, 0), (1, 1, 1)\}$\\
 \cline{2-2}
 &  $ {\rm SR}: \{ x_ 3  x_ 4, x_ 6  x_ 7, x_ 1  x_ 2  x_ 5 \}$\\
 & $ \sigma: \{x_3 \leftrightarrow x_6\, \& \,  \, x_4 \leftrightarrow x_7\} $ , \, \, \, \, \, $x_i \equiv K3 \, \, \forall i = \{2,..., 7\}$\\
& KC:\, $\{(1, 0, 0), (0, 0, 1), (1, 1, 1)\}$\\
\hline
& ${\rm GLSM}: [ (2, 0, 2), (0, 0, 1), (1, -1, 1), (1, -1, 1), (0, 0, 1), (0, 2, 0), (2, 0, 0) ]$ \\
\cline{2-2}
   \multirow{2} {*} {9} &  $ {\rm SR}: \{ x_ 2  x_ 5, x_ 3  x_ 4, x_ 1  x_ 6  x_ 7 \}$\\
 & $ \sigma_1: \{x_2 \leftrightarrow x_3\, \&  \, x_4 \leftrightarrow x_5\equiv K3\} $; $ \sigma_2: x_6 \leftrightarrow x_7 = \{1,4,0,2\}$; $\sigma_3 = \sigma_1 \cup \sigma_2$ \\
& KC: \, $\{(0, 0, 1), (1, -1, 1), (2, 0, 2)\}$\\
 \hline
    & ${\rm GLSM}: [(0, 0, 1), (1, 2, -1), (1, 2, -1), (0, 0, 1), (0, 1, 0), (0, 1, 0), (1, 0, 0) ]$\\
\cline{2-2}
\multirow{2} {*} {10} &  $ {\rm SR}: \{ x_ 1  x_ 4, x_ 5  x_ 6, x_ 2  x_ 3  x_ 7 \}$\\
 & $ \sigma: \{x_1 \leftrightarrow x_5\, \& \, x_4 \leftrightarrow x_6\equiv K3\} $; \, KC: \, $\{(0, 1, 0), (0, 0, 1), (1, 2, 0)\}$ \\
\cline{2-2}
\multirow{2} {*} {11} &  $ {\rm SR}: \{ x_ 2  x_ 3, x_ 5  x_ 6, x_ 1  x_ 4  x_ 7 \}$\\
 & $ \sigma: \{x_2 \leftrightarrow x_5\, \&  \, x_3 \leftrightarrow x_6\equiv K3\} $; \, KC: $\{(0, 1, 0), (1, 2, -1), (1, 2, 0)\}$\\
 \hline
 & ${\rm GLSM}: [(1, 0, 1), (0, 1, 1), (-1, -1, 1), (0, 0, 1), (0, 0, 1), (0, 1, 0), (1, 0, 0)]$ \\
\cline{2-2}
   \multirow{3} {*} {12} &  $ {\rm SR}: \{ x_ 1  x_ 7, x_ 2  x_ 6, x_ 3  x_ 4  x_ 5 \}$\, \, \, \, \, \,  $x_3 \equiv dP_9$ \\
 & $ \sigma: \{x_1 \leftrightarrow x_2 \equiv \{1,0,3,37\}\, \& \, x_6 \leftrightarrow x_7 \equiv dP_8\}$  \\
& KC:\, $\{(0, 1, 1), (1, 0, 1), (0, 0, 1)\}$\\
 \hline& ${\rm GLSM}:[(-1, 1, 1), (0, 0, 1), (2, -1, 0), (0, 1, 0), (1, 0, 0), (1, 0, 0), (1, 0, 0)] $\\
\cline{2-2}
   \multirow{2} {*} {13} &  $ {\rm SR}: \{ x_ 1  x_ 2, x_ 1  x_ 4, x_ 2  x_ 3, x_ 3  x_ 5  x_ 6  x_ 7, x_ 4  x_ 5  x_ 6  x_ 7\}$ \\
 & $ \sigma: \{x_1 \leftrightarrow x_2 \equiv dP_7 \, \& \, x_3 \leftrightarrow x_4 \equiv dP_{20} \}$; \, KC:$\{(1, 0, 0), (0, 1, 2), (1, 0, 1), (0, 1, 1)\}$ \\
  \hline
& ${\rm GLSM}: [(1, 2, -1), (1, 2, 0), (0, 0, 1), (0, 0, 1), (0, 1, 0), (0, 1, 0), (1, 0, 0)]$\\
\cline{2-2}
   \multirow{2} {*} {14} &  $ {\rm SR}: \{  x_ 3  x_ 4, x_ 5  x_ 6, x_ 1  x_ 2  x_ 7 \} $\\
 &  $ \sigma: \{x_3 \leftrightarrow x_5\, \&  \, x_4 \leftrightarrow x_6\equiv K3\} $; \, KC: \, $\{(0, 1, 0), (0, 0, 1), (1, 2, 0)\}$ \\
\hline&  ${\rm GLSM}: [(-1, 1, 1), (0, 0, 1), (1, 1, 0), (0, 1, 0), (0, 1, 0), (1, 0, 0), (1, 0, 0)] $ \\
\cline{2-2}
   \multirow{2} {*} {15} &  $ {\rm SR}: \{ x_ 1  x_ 2, x_ 1  x_ 4  x_ 5, x_ 3  x_ 4  x_ 5, x_ 3  x_ 6  x_ 7, x_ 2  x_ 6  x_ 7 \}$\\
 & $ \sigma: x_1 \leftrightarrow x_2 \equiv dP_{11}$; \, KC:\, $\{(1, 1, 0), (1, 1, 1), (0, 1, 1), (0, 2, 1)\}$ \\
\hline
  \end{tabular}
\end{table}

\begin{table}[H]
  \centering
  \begin{tabular}{|c||c|}
  \hline
   No. & Toric data   \\
    \hline \hline
    & ${\rm GLSM}: [(-1, 1, 1), (0, 1, 0), (0, 0, 1), (0, 0, 1), (0, 1, 0), (1, 0, 0), (1, 0, 0)] $\\
\cline{2-2}
   \multirow{3} {*} {16} &  $ {\rm SR}: \{ x_ 2  x_ 5, x_ 6  x_ 7, x_ 1  x_ 3  x_ 4 \}$\\
 & $ \sigma_1: \{x_2 \leftrightarrow x_6\, \&  \, x_5 \leftrightarrow x_7\equiv K3\}$, \, {\rm and}  $x_1 \equiv dP_{19}$ \\
& KC: \, $\{(0, 1, 0), (1, 0, 0), (0, 1, 1)\}$\\
\cline{2-2}
\multirow{3} {*} {17} &  $ {\rm SR}: \{  x_ 3  x_ 4, x_ 6  x_ 7, x_ 1  x_ 2  x_ 5 \}$\\
 & $ \sigma_1: \{x_3 \leftrightarrow x_6\, \&  \, x_4 \leftrightarrow x_7\equiv K3\}$, \, {\rm and}  $x_1 \equiv dP_{19} $ \\
& KC: \, $\{(0, 0, 1), (1, 0, 0), (0, 1, 1)\}$\\
 \hline
 & ${\rm GLSM}: [(-1, 1, 0), (-1, 0, 1), (0, 0, 1), (0, 1, 0), (2, 0, 0), (1, 0, 0), (1, 0, 0)]$ \\
\cline{2-2}
   \multirow{2} {*} {18} &  $ {\rm SR}: \{  x_ 1  x_ 4, x_ 2  x_ 3, x_ 5  x_ 6  x_ 7 \}$\\
 & $ \sigma:\{x_1 \leftrightarrow x_2 \equiv dP_{8}\, \& \, x_3 \leftrightarrow x_4 \equiv \{1,0,2,29\},\, x_1 \leftrightarrow x_2\} $ \\
& KC:\, $\{(0, 0, 1), (0, 1, 0), (1, 0, 0)\}$\\
 \hline
    & ${\rm GLSM}: [ (-1, 1, 0), (-1, 0, 1), (0, 0, 1), (0, 0, 1), (0, 1, 0), (0, 1, 0), (1, 0, 0) ]$\\
\cline{2-2}
   \multirow{2} {*} {19} &  $ {\rm SR}: \{ x_ 1  x_ 2, x_ 1  x_ 5  x_ 6, x_ 2  x_ 3  x_ 4, x_ 3  x_ 4  x_ 7, x_ 5  x_ 6  x_ 7 \}$\\
 & $ \sigma:\{x_1 \leftrightarrow x_2 \equiv dP_{10} \& \, x_3\leftrightarrow x_4\, \& \, x_5 \leftrightarrow x_6\equiv {\rm SD2}\} $ \\
& KC: \, $\{(0, 0, 1), (0, 1, 0), (-1, 1, 1)\}$\\
 \hline
& ${\rm GLSM}: [ (-1, 0, 1), (0, -1, 1), (0, 0, 1), (0, 1, 0), (1, 0, 0), (0, 1, 0), (1, 0, 0) ]$\\
\cline{2-2}
   \multirow{2} {*} {20} &  $ {\rm SR}: \{ x_ 4  x_ 6, x_ 5  x_ 7, x_ 1  x_ 2  x_ 3 \}$\\
 & $ \sigma_1: x_1 \leftrightarrow x_2 \equiv dP_{16}$; \,  $ \sigma_2: \{x_4\leftrightarrow x_5\, \& \,  x_6 \leftrightarrow x_7\equiv K3\}$; $\sigma_3=\sigma_1 \cup \sigma_2$ \\
& KC:\, $\{(0, 1, 0), (1, 0, 0), (0, 0, 1)\}$\\
 \hline
& ${\rm GLSM}: [(1, 0, 1), (1, 1, 0), (0, 0, 1), (0, 0, 1), (0, 1, 0), (0, 1, 0), (1, 0, 0) ]$ \\
\cline{2-2}
  \multirow{3} {*} {21} &  $ {\rm SR}: \{ x_ 1  x_ 7, x_ 2  x_ 7, x_ 1  x_ 3  x_ 4, x_ 2  x_ 5  x_ 6, x_ 3  x_ 4  x_ 5  x_ 6 \}$ \\
& $\sigma: \{x_1 \leftrightarrow x_2 \equiv {\rm SD2}\, \& \, x_3 \leftrightarrow x_5\, \& \, x_4 \leftrightarrow x_6 \equiv \{1,0,1,23\}\}$ \, \, \, \, $x_7 \equiv dP_6$ \\
& KC: \, $\{(1, 0, 1), (1, 1, 0), (1, 1, 1)\}$\\
\cline{2-2}
  \multirow{3} {*} {22} &  $ {\rm SR}: \{ x_ 3  x_ 4, x_ 5  x_ 6, x_ 1  x_ 2  x_ 7 \}$ \\
 &  $ \sigma_1: \{x_1 \leftrightarrow x_2 \equiv {\rm SD2}\}$; \, $ \sigma_2:\{x_3 \leftrightarrow x_5\, \&  \, x_4 \leftrightarrow x_6\equiv K3 \}; $ \, $\sigma_3 = \sigma_1 \cup \sigma_2 $\\
& KC: \, $\{(0, 0, 1), (0, 1, 0), (1, 1, 1)\}$\\
 \hline
& ${\rm GLSM}: [(0, 0, 1), (0, 0, 1), (1, 2, 0), (1, 2, 0), (0, 1, 0), (0, 1, 0), (1, 0, 0) ] $\\
\cline{2-2}
   \multirow{2} {*} {23} &  $ {\rm SR}: \{ x_ 1  x_ 2, x_ 5  x_ 6, x_ 3  x_ 4  x_ 7 \}$\\
 & $ \sigma: \{x_1 \leftrightarrow x_5\, \&  \, x_2 \leftrightarrow x_6\equiv K3 \} $\,  KC: $\{(0, 0, 1), (0, 1, 0), (1, 2, 0)\}$\\
 \hline
& ${\rm GLSM}: [(0, 0, 1), (0, 0, 1), (1, 1, 0), (0, 1, 0), (0, 1, 0), (1, 0, 0), (1, 0, 0) ]$\\
\cline{2-2}
   \multirow{2} {*} {24} &  $ {\rm SR}: \{ x_ 1  x_ 2, x_ 4  x_ 5, x_ 3  x_ 6  x_ 7 \}$\\
 & $ \sigma: \{x_1 \leftrightarrow x_4\, \&  \, x_2\leftrightarrow x_5\equiv K3 \} $,\, KC: $\{(0, 0, 1), (0, 1, 0), (1, 1, 0)\}$\\
\cline{2-2}
\multirow{2} {*} {25} &  $ {\rm SR}: \{ x_ 1  x_ 2, x_ 6  x_ 7, x_ 3  x_ 4  x_ 5  \}$\\
 & $ \sigma: \{x_1 \leftrightarrow x_6\, \&  \, x_2\leftrightarrow x_7\equiv K3 \} $,\, KC: $\{(0, 0, 1), (1, 0, 0), (1, 1, 0)\}$\\
 \hline
     & ${\rm GLSM}: [(0, 0, 1), (0, 0, 1), (-1, 1, 0), (0, 1, 0), (0, 1, 0), (1, 0, 0), (1, 0, 0) ]$\\
\cline{2-2}
   \multirow{2} {*} {26} &  $ {\rm SR}: \{ x_ 1  x_ 2, x_ 6  x_ 7, x_ 3  x_ 4  x_ 5 \}$\\
 & $ \sigma: \{x_1 \leftrightarrow x_6\, \& \, x_2\leftrightarrow x_7\equiv K3 \} $  \, {\rm and}  $x_3\equiv dP_{13}$\\
& KC: $\{(0, 0, 1), (0, 1, 0), (1, 0, 0)\}$\\
 \hline
& ${\rm GLSM}: [(0, 1, 0), (0, 0, 1), (0, 0, 1), (0, 1, 0), (1, 0, 0), (1, 0, 0), (1, 0, 0) ]$ \\
\cline{2-2}
   \multirow{2} {*} {27} &  $ {\rm SR}: \{ x_ 1  x_ 4, x_ 2  x_ 3, x_ 5  x_ 6  x_ 7 \}$\\
 & $ \sigma:  \{x_1 \leftrightarrow x_2 \,  \& \, x_3 \leftrightarrow x_4\equiv K3 \}$ \, {\rm and}  $x_5= x_6=x_7 \equiv {\rm SD2}$\\
& KC: $\{(1, 0, 0), (0, 0, 1), (0, 1, 0)\}$\\
 \hline
 \end{tabular}
\end{table}

\begin{table}[H]
  \centering
  \begin{tabular}{|c||c|}
  \hline
   No. & Toric data   \\
    \hline \hline
    & ${\rm GLSM}: [(0, 1, 1), (-1, 1, 0), (0, 0, 1), (0, 0, 1), (0, 1, 0), (1, 0, 0), (1, 0, 0)]$ \\
\cline{2-2}
   \multirow{2} {*} {28} &  $ {\rm SR}: \{ x_ 3  x_ 4, x_ 6  x_ 7, x_ 1  x_ 2  x_ 5 \}$\\
 & $ \sigma: \{x_3 \leftrightarrow x_6\, \& \, x_4 \leftrightarrow x_7\equiv K3 \} $ \, {\rm and }\, $x_2 \equiv dP_{10} $ \\
& KC: $\{(1, 0, 0), (0, 0, 1), (0, 1, 1)\}$\\
 \hline
& ${\rm GLSM}: [(-1, 1, 0), (-1, 0, 1), (0, 0, 1), (0, 1, 0), (1, 0, 0), (1, 0, 0), (1, 0, 0) ]$\\
\cline{2-2}
   \multirow{2} {*} {29} &  $ {\rm SR}: \{ x_ 1  x_ 4, x_ 2  x_ 3, x_ 5  x_ 6  x_ 7 \}$\\
 & $ \sigma: \{x_1 \leftrightarrow x_2 \equiv dP_7\, \& \, x_3 \leftrightarrow x_4 \equiv \{1,0,3,38\} \}$ \, \, \, $x_5 = x_6 = x_7 \equiv SD2$ \\
 & KC: $\{(0, 0, 1), (0, 1, 0), (1, 0, 0)\}$\\
\hline
& ${\rm GLSM}: [(1, 2, 1), (1, 2, 0), (0, 0, 1), (0, 0, 1), (0, 1, 0), (0, 1, 0), (1, 0, 0)]$\\
\cline{2-2}
   \multirow{3} {*} {30} &  $ {\rm SR}: \{ x_ 3  x_ 4, x_ 5  x_ 6, x_ 1  x_ 2  x_ 7 \}$\\
 & $ \sigma: \{x_3 \leftrightarrow x_5\, \& \, x_4 \leftrightarrow x_6 \equiv K3 \} $ \, {\rm and }\, $x_7 \equiv W $ \\
& KC: $\{(0, 1, 0), (0, 0, 1), (1, 2, 1)\}$\\
 \hline
& ${\rm GLSM}: [(-1, -1, 1), (0, 0, 1), (0, 0, 1), (0, 1, 0), (0, 1, 0), (1, 0, 0), (1, 0, 0)]$\\
\cline{2-2}
   \multirow{2} {*} {31} &  $ {\rm SR}: \{ x_ 4  x_ 5, x_ 6  x_ 7, x_ 1  x_ 2  x_ 3 \}$\\
& $ \sigma: \{x_4 \leftrightarrow x_6\, \& \, x_5 \leftrightarrow x_7\equiv K3 \} $ \, {\rm and }\, $x_1 \equiv dP_7$ \\
& KC: $\{(1, 0, 0), (0, 1, 0), (0, 0, 1)\}$\\
 \hline
& ${\rm GLSM}: [(1, 1, 1), (0, 1, 0), (0, 0, 1), (0, 0, 1), (0, 1, 0), (1, 0, 0), (1, 0, 0) ]$ \\
\cline{2-2}
   \multirow{9} {*} {32-34} &  $ {\rm SR}: \{ x_ 2  x_ 5, x_ 3  x_ 4, x_ 1  x_ 6  x_ 7 \}$\\
   & $ \sigma: \{x_2 \leftrightarrow x_3 \,\&\,  x_4 \leftrightarrow x_5 \}$; \,\,\,  $x_i \equiv K3 \, \, \forall i = \{2,..., 7\}$\\
& KC: $\{(0, 0, 1), (0, 1, 0), (1, 1, 1)\}$\\
   \cline{2-2}
 &  $ {\rm SR}: \{ x_ 2  x_ 5, x_ 6  x_ 7, x_ 1  x_ 3  x_ 4 \}$\\
 & $ \sigma:   \{x_2 \leftrightarrow x_6, \,\&\,  x_5 \leftrightarrow x_7\} $, \, \, \,  $x_i \equiv K3 \, \, \forall i = \{2,..., 7\}$\\
& KC: $\{(1, 0, 0), (0, 1, 0), (1, 1, 1)\}$\\
   \cline{2-2}
 &  $ {\rm SR}: \{ x_ 3  x_ 4, x_ 6  x_ 7, x_ 1  x_ 2  x_ 5 \}$\\
 &  $ \sigma: \{x_3 \leftrightarrow x_6,  \,\&\, x_4 \leftrightarrow x_7\} $, \, \, \,  $x_i \equiv K3 \, \, \forall i = \{2,..., 7\}$\\
& KC: $\{(1, 0, 0), (0, 0, 1), (1, 1, 1)\}$\\
 \hline
& ${\rm GLSM}: [(0, 4, 4), (0, 0, 1), (-1, 1, 1), (-1, 1, 1), (0, 0, 1), (0, 2, 0), (2, 0, 0)]$ \\
\cline{2-2}
   \multirow{3} {*} {35} &  $ {\rm SR}: \{ x_ 2  x_ 5, x_ 3  x_ 4, x_ 1  x_ 6  x_ 7 \}$\\
 & $ \sigma_1: x_6 \leftrightarrow x_7 \equiv \{1,3,0,2\}$; $ \sigma_2: \{x_2 \leftrightarrow x_3 \, \&  \, x_4 \leftrightarrow x_5 \equiv K3\}$,$\sigma_3 = \sigma_1 \cup \sigma_2$ \\
& KC: $\{(-1, 1, 1), (0, 0, 1), (0, 2, 2)\}$\\
 \hline
& ${\rm GLSM}: [(1, 2, 1), (1, 2, 1), (0, 0, 1), (0, 0, 1), (0, 1, 0), (0, 1, 0), (1, 0, 0)]$ \\
\cline{2-2}
   \multirow{2} {*} {36} &  $ {\rm SR}: \{ x_ 3  x_ 4, x_ 5  x_ 6, x_ 1  x_ 2  x_ 7 \}$\\
 & $ \sigma: \{x_3 \leftrightarrow x_5\, \& \, x_4 \leftrightarrow x_6\equiv K3 \} $ \, {\rm and }\, $x_7 \equiv dP_{1} $ \\
& KC: $\{(0, 1, 0), (0, 0, 1), (1, 2, 1)\}$\\
 \hline
 & ${\rm GLSM}: [(2, 4, 1), (1, 2, -1), (0, 0, 1), (0, 0, 1), (0, 1, 0), (0, 1, 0), (1, 0, 0)]$ \\
\cline{2-2}
   \multirow{3} {*} {37} &  $ {\rm SR}: \{ x_ 3  x_ 4, x_ 5  x_ 6, x_ 1  x_ 2  x_ 7 \}$\\
 & $ \sigma: \{x_3 \leftrightarrow x_5\, \& \, x_4 \leftrightarrow x_6 \equiv K3 \} $ \\
& KC: $\{(0, 1, 0), (0, 0, 1), (2, 4, 1)\}$\\
 \hline
 & ${\rm GLSM}: [(1, 3, 2), (-1, 1, 1), (0, 0, 1), (0, 1, 0), (0, 1, 0), (1, 0, 0), (1, 0, 0) ]$ \\
\cline{2-2}
   \multirow{2} {*} {38} &  $ {\rm SR}: \{ x_ 2  x_ 3, x_ 1  x_ 4  x_ 5, x_ 1  x_ 6  x_ 7, x_ 2  x_ 4  x_ 5, x_ 3  x_ 6  x_ 7 \}$\\
 & $ \sigma:\{x_2 \leftrightarrow x_3 \equiv dP_{13}\, \& \, x_4\leftrightarrow x_6\equiv K3\, \& \, x_5 \leftrightarrow x_7\equiv K3\} $ \\
& KC: $\{(1, 3, 2), (0, 1, 1), (1, 3, 3), (0, 4, 3)\}$\\
 \hline
\end{tabular}
\end{table}
\begin{table}[H]
  \centering
  \begin{tabular}{|c||c|}
  \hline
   No. & Toric data   \\
    \hline \hline
    & ${\rm GLSM}: [(1,1,2), (-1,-1, 1), (0, 0, 1), (0, 1,0), (0, 1, 0), (1, 0, 0), (1, 0, 0)]$ \\
\multirow{3} {*} {43} &  $ {\rm SR}: \{ x_ 4  x_ 5, x_ 6  x_ 7, x_ 1  x_ 2  x_ 3 \}$\\
 & $ \sigma: \{x_4 \leftrightarrow x_6 \, \&  \, x_5 \leftrightarrow x_7 \equiv K3 \} $  \\
& KC: $\{(1, 0, 0), (0, 1, 0), (1, 1, 2)\}$\\
\hline
& ${\rm GLSM}: [(2, 4, 3), (1, 2, 1), (0, 0, 1), (0, 0, 1), (0, 1, 0), (0, 1, 0), (1, 0, 0)]$ \\
\multirow{3} {*} {44} &  $ {\rm SR}: \{ x_ 3  x_ 4, x_ 5  x_ 6, x_ 1  x_ 2  x_ 7 \}$\\
 & $ \sigma: \{x_3 \leftrightarrow x_5 \, \&  \, x_4 \leftrightarrow x_6 \equiv K3 \} $  \\
& KC: $\{(0, 1, 0), (0, 0, 1), (2, 4, 3)\}$\\
\hline
& ${\rm GLSM}: [(3, 6, 6), (2, 4, 4), (0, 0, 1), (0, 1, 0), (0, 0, 1), (0, 1, 0), (1, 0, 0)]$ \\
\cline{2-2}
   \multirow{2} {*} {45} &  $ {\rm SR}: \{ x_ 3  x_ 5, x_ 4  x_ 6, x_ 1  x_ 2  x_ 7 \}$\\
 & $ \sigma: \{x_3 \leftrightarrow x_4 \, \&  \, x_5 \leftrightarrow x_6 \equiv K3 \} $ \, {\rm and }\, $x_7 \equiv dP_{1}$ \\
& KC: $\{(0, 1, 0), (0, 0, 1), (1, 2, 2)\}$\\
 \hline
\end{tabular}
\caption{List of $CY_3$ spaces with $h^{1,1}=3$ for the possibility of $h^{1,1}_-(CY_3/\sigma) \neq 0$
under divisor exchange involutions.
}
\label{GLSMlist}
\end{table}

\section{List of Non-toric Spaces}
\label{sec:non-toric}
The spaces presented in this section are special examples in which the reflection involution can also
result in $h^{1,1}_-(CY_3/ \sigma)\neq 0$. The special thing about these spaces is
the fact that in each case, a single divisor itself has several disjoint ${\mathbb P}^2$s or
two disjoint $dP_1$s, and under some reflection involution $x_i \rightarrow -x_i$, these
are exchanged within the divisor itself.
\begin{table}[H]
  \centering
  \begin{tabular}{|c||c|}
  \hline
   No. & Toric data   \\
    \hline \hline
  & ${\rm GLSM}: [(2, 6), (1, 3), (0, 1), (0, 1), (0, 1), (1, 0) ] $\\
  \cline{2-2}
1  &  $ {\rm SR}: \{ x_ 1  x_ 2 x_ 6,  x_ 3  x_ 4  x_ 5 \}$,\, $x_6 \equiv \IP^2 \sqcup \IP^2 $  \\
  & $ \sigma: x_1 \leftrightarrow -x_1 \, , $ \, $h^{1,1} = 2_+ +1_-$ \\
   \hline
& ${\rm GLSM}: [(2, 4, 4), (1, 2, 2), (0, 0, 1), (0, 1, 0), (0, 0, 1), (0, 1, 0), (1, 0, 0) ] $\\
\cline{2-2}
2   &  $ {\rm SR}: \{ x_ 3  x_ 5, x_ 4  x_ 6, x_ 1  x_ 2  x_ 7 \}$,\,  $x_7 \equiv dP_1 \sqcup dP_1$\\
   & $ \sigma: x_1 \leftrightarrow -x_1 \, , $ \,$h^{1,1} = 3_+ +1_-$\\
 \hline
& ${\rm GLSM}: [(2, 4, 8), (1, 2, 4), (0, 1, 2), (0,0,1), (0, 0, 1), (0, 1, 0), (1, 0, 0) ] $\\
\cline{2-2}
3&  $ {\rm SR}: \{ x_ 3  x_ 6, x_ 4  x_ 5, x_ 1  x_ 2  x_ 7 \}$,\,  $x_7 \equiv dP_1 \sqcup dP_1$\\
   & $ \sigma: x_1 \leftrightarrow -x_1 \, , $ \,$h^{1,1} = 3_+ +1_-$\\
   \hline
 \multirow{4} {*} {4}  & ${\rm GLSM}: [(1, 3), (1, 3), (0, 1), (0, 1), (0, 1), (1, 0) ] $\\
  \cline{2-2}
  &  ${\rm SR}:\{x_1 x_2 x_6, x_3 x_4 x_5 \}  $,\, $x_6 \equiv \IP^2 \sqcup \IP^2 \sqcup \IP^2$ \\
   & $ \sigma_1: x_1 \leftrightarrow -x_1 \,\, {\rm or} $ \,$ \sigma_2: x_2 \leftrightarrow -x_2 \, , $ \,$h^{1,1} = 3_+ +1_-$\\
   & $ \sigma_3: \{x_1 \leftrightarrow -x_1 \,\&\,   x_2 \leftrightarrow -x_2 \}$,\, $\,h^{1,1} = 2_+ +2_-$\\
   \hline
    & ${\rm GLSM}: [(-1,3,9),(-1,1,3),(0,0,2),(0,0,2),(0,0,2),(0,2,0),(2,0,0) ] $\\
    \cline{2-2}
  5 & ${\rm SR}: \{x_1 x_2, x_6 x_7, x_3 x_4 x_5\}$, \, $x_6 \equiv \IP^2 \sqcup \IP^2$\\
 & $ \sigma: x_1 \leftrightarrow -x_1 \, , $ \,$h^{1,1} = 3_+ +1_-$\\
   \hline
\end{tabular}
\end{table}

\begin{table}[H]
  \centering
\begin{tabular}{|c||c|}
   \hline
 \multirow{5} {*} {6-7}   & ${\rm GLSM}: [(1,3,2),(0,1,1),(0,1,1),(0,0,1),(0,0,1),(0,1,0),(1,0,0) ] $\\
       \cline{2-2}
 & ${\rm SR}: \{x_1 x_6, x_1 x_7, x_2 x_3 x_5, x_4 x_5 x_7, x_2 x_3 x_4 x_5\}$, \, $x_7 \equiv dP_1 \sqcup dP_1$\\
 & $ \sigma_1: x_4 \leftrightarrow -x_4 \, , $ \,$ \sigma_2: x_5 \leftrightarrow -x_5 \, , $ \,$h^{1,1} = 3_+ +1_-$\\
\cline{2-2}
 & ${\rm SR}: \{x_1 x_7, x_4 x_5, x_2 x_3 x_6\}$, \, $x_7 \equiv \IP^2 \sqcup \IP^2$\\
  & $ \sigma_1: x_4 \leftrightarrow -x_4 \, , $ \,$ \sigma_2: x_5 \leftrightarrow -x_5 \, , $ \,$h^{1,1} = 3_+ +1_-$\\
   \hline
       & ${\rm GLSM}: [(1,2,1),(0,-1,1),(0,-1,1),(0,0,1),(0,1,0),(0,1,0),(1,0,0) ] $\\
       \cline{2-2}
  8 & ${\rm SR}: \{x_1 x_7, x_1 x_5 x_6, x_2 x_3 x_4, x_2 x_3 x_7, x_4 x_5 x_6\}$, \, $x_7 \equiv \IP^2 \sqcup \IP^2$\\
 & $ \sigma_1: x_2 \leftrightarrow -x_2 \, , $ \,$ \sigma_2: x_3 \leftrightarrow -x_3 \, , $ \,$h^{1,1} = 3_+ +1_-$\\
   \hline
       & ${\rm GLSM}: [(1,3,0),(0,0,1),(0,0,1),(0,1,0),(0,1,0),(0,1,0),(1,0,0)] $\\
       \cline{2-2}
  9 & ${\rm SR}: \{x_1 x_7, x_2 x_3, x_4 x_5 x_6\}$, \, $x_7 \equiv \IP^2 \sqcup \IP^2$\\
 & $ \sigma_1: x_2 \leftrightarrow -x_2 \, , $ \,$ \sigma_2: x_3 \leftrightarrow -x_3 \, , $ \,$h^{1,1} = 3_+ +1_-$\\
   \hline
  \multirow{3} {*} {10}  & ${\rm GLSM}: [(1,3,1),(1,3,-1),(0,0,1),(0,1,0),(0,1,0),(0,1,0),(1,0,0) ] $\\
   \cline{2-2}
   & ${\rm SR}: \{x_1 x_3, x_2 x_7, x_4 x_5 x_6\}$, \, $x_7 \equiv \IP^2 \sqcup \IP^2$\\
 & $ \sigma: x_1 \leftrightarrow -x_1 \, , $  \,$h^{1,1} = 3_+ +1_-$\\
 \hline
\multirow{3} {*} {11}  & ${\rm GLSM}: [(-1,-3,1),(0,0,1),(1,3,0),(0,1,0),(0,1,0),(0,1,0),(1,0,0) ] $\\
   \cline{2-2}
   & ${\rm SR}: \{x_1 x_2, x_3 x_7, x_4 x_5 x_6\}$, \, $x_7 \equiv \IP^2 \sqcup \IP^2$\\
 & $ \sigma_1: x_1 \leftrightarrow -x_1 \,  $ , $\sigma_2: x_2 \leftrightarrow -x_2$, \,$h^{1,1} = 3_+ +1_-$\\
  \hline
    \multirow{5} {*} {12-13}    & ${\rm GLSM}: [(2,1,5),(1,1,2),(0,-1,1),(0,0,1),(0,0,1),(0,1,0),(1,0,0) ] $\\
       \cline{2-2}
  & ${\rm SR}: \{x_2 x_6, x_1 x_2 x_7, x_1 x_3 x_7, x_3 x_4 x_5, x_4 x_5 x_6\}$, \, $x_7 \equiv \IP^2 \sqcup \IP^2$\\
 & $ \sigma_1: x_1 \leftrightarrow -x_1 \, , $ \,$ \sigma_2: x_3 \leftrightarrow -x_3 \, , $ \,$h^{1,1} = 3_+ +1_-$\\
\cline{2-2}
  & ${\rm SR}: \{x_3 x_7, x_1 x_2 x_6, x_1 x_2 x_7, x_3 x_4 x_5 x_4 x_5 x_6\}$, \, $x_7 \equiv \IP^2 \sqcup \IP^2$\\
 & $ \sigma_1: x_1 \leftrightarrow -x_1 \, , $ \,$ \sigma_2: x_3 \leftrightarrow -x_3 \, , $ \,$h^{1,1} = 3_+ +1_-$\\
   \hline
    \multirow{5} {*} {14-15}       & ${\rm GLSM}: [(2,3,3),(1,2,1),(0,-1,1),(0,0,1),(0,1,0),(0,1,0),(1,0,0) ] $\\
       \cline{2-2}
 & ${\rm SR}: \{x_3 x_4, x_1 x_2 x_7, x_1 x_3 x_7, x_2 x_5 x_6, x_4 x_5 x_6\}$, \, $x_7 \equiv \IP^2 \sqcup \IP^2$\\
 & $ \sigma_1: x_1 \leftrightarrow -x_1 \, , $ \,$ \sigma_2: x_3 \leftrightarrow -x_3 \, , $ \,$h^{1,1} = 3_+ +1_-$\\
\cline{2-2}
 & ${\rm SR}: \{x_2 x_7, x_1 x_3 x_4,x_1 x_3 x_7, x_2 x_5 x_6, x_4 x_5 x_6\}$, \, $x_7 \equiv \IP^2 \sqcup \IP^2$\\
 & $ \sigma_1: x_1 \leftrightarrow -x_1 \, , $ \,$ \sigma_2: x_3 \leftrightarrow -x_3 \, , $ \,$h^{1,1} = 3_+ +1_-$\\
   \hline
    \multirow{5} {*} {16-17}       & ${\rm GLSM}: [(2,6,4),(1,3,2),(0,1,1),(0,0,1),(0,1,0),(0,1,0),(1,0,0) ] $\\
    \cline{2-2}
 & ${\rm SR}: \{x_3 x_4, x_4 x_7, x_1 x_2 x_7, x_3 x_5 x_6 x_1 x_2 x_5 x_6\}$, \, $x_7 \equiv \IP^2 \sqcup \IP^2$\\
 & $ \sigma: x_1 \leftrightarrow -x_1 \, , $  \,$h^{1,1} = 3_+ +1_-$\\
\cline{2-2}
 & ${\rm SR}: \{x_3 x_4, x_5 x_6, x_1 x_2 x_7\}$, \, $x_7 \equiv dP_1 \sqcup dP_1$\\
 & $ \sigma: x_1 \leftrightarrow -x_1 \, , $ \,$h^{1,1} = 3_+ +1_-$\\
   \hline
\end{tabular}
\caption{List of {\it non-toric} $CY_3$ spaces with the possibility of $h^{1,1}_-(CY_3/\sigma) \neq 0$
with reflection involutions. In all these examples, $h^{1,1}(CY_3)$ is more
than the equivalence relations for GLSM charges, and hence one K\"ahler deformation is {\it non-toric}.}
\label{GLSMlist2}
\end{table}

\clearpage
\nocite{*}
\bibliography{ClassifyingCY3}

\providecommand{\href}[2]{#2}\begingroup\raggedright\begin{thebibliography}{10}

\bibitem{Grana:2005jc}
M.~Grana, ``{Flux compactifications in string theory: A Comprehensive
  review},'' {\em Phys.Rept.} {\bf 423} (2006) 91--158,
\href{http://www.arXiv.org/abs/hep-th/0509003}{{\tt hep-th/0509003}}.

\bibitem{Lust:2006zg}
D.~Lust, S.~Reffert, E.~Scheidegger, W.~Schulgin, and S.~Stieberger, ``{Moduli
  Stabilization in Type IIB Orientifolds (II)},'' {\em Nucl.Phys.} {\bf B766}
  (2007) 178--231,
\href{http://www.arXiv.org/abs/hep-th/0609013}{{\tt hep-th/0609013}}.

\bibitem{Kachru:2003aw}
S.~Kachru, R.~Kallosh, A.~D. Linde, and S.~P. Trivedi, ``{De Sitter vacua in
  string theory},'' {\em Phys.Rev.} {\bf D68} (2003) 046005,
\href{http://www.arXiv.org/abs/hep-th/0301240}{{\tt hep-th/0301240}}.

\bibitem{Balasubramanian:2005zx}
V.~Balasubramanian, P.~Berglund, J.~P. Conlon, and F.~Quevedo, ``{Systematics
  of moduli stabilisation in Calabi-Yau flux compactifications},'' {\em JHEP}
  {\bf 0503} (2005) 007,
\href{http://www.arXiv.org/abs/hep-th/0502058}{{\tt hep-th/0502058}}.

\bibitem{Gukov:1999ya}
S.~Gukov, C.~Vafa, and E.~Witten, ``{CFT's from Calabi-Yau four folds},'' {\em
  Nucl.Phys.} {\bf B584} (2000) 69--108,
\href{http://www.arXiv.org/abs/hep-th/9906070}{{\tt hep-th/9906070}}.

\bibitem{Taylor:1999ii}
T.~R. Taylor and C.~Vafa, ``{R R flux on Calabi-Yau and partial supersymmetry
  breaking},'' {\em Phys.Lett.} {\bf B474} (2000) 130--137,
\href{http://www.arXiv.org/abs/hep-th/9912152}{{\tt hep-th/9912152}}.

\bibitem{Dasgupta:1999ss}
K.~Dasgupta, G.~Rajesh, and S.~Sethi, ``{M theory, orientifolds and G -
  flux},'' {\em JHEP} {\bf 9908} (1999) 023,
\href{http://www.arXiv.org/abs/hep-th/9908088}{{\tt hep-th/9908088}}.

\bibitem{Witten:1996bn}
E.~Witten, ``{Non-Perturbative Superpotentials In String Theory},'' {\em Nucl.
  Phys.} {\bf B474} (1996) 343--360,
\href{http://www.arXiv.org/abs/hep-th/9604030}{{\tt hep-th/9604030}}.

\bibitem{Becker:2002nn}
K.~Becker, M.~Becker, M.~Haack, and J.~Louis, ``{Supersymmetry breaking and
  alpha-prime corrections to flux induced potentials},'' {\em JHEP} {\bf 0206}
  (2002) 060,
\href{http://www.arXiv.org/abs/hep-th/0204254}{{\tt hep-th/0204254}}.

\bibitem{Blumenhagen:2007sm}
R.~Blumenhagen, S.~Moster, and E.~Plauschinn, ``{Moduli Stabilisation versus
  Chirality for MSSM like Type IIB Orientifolds},'' {\em JHEP} {\bf 0801}
  (2008) 058,
\href{http://www.arXiv.org/abs/0711.3389}{{\tt 0711.3389}}.

\bibitem{Collinucci:2008sq}
A.~Collinucci, M.~Kreuzer, C.~Mayrhofer, and N.-O. Walliser, ``{Four-modulus
  'Swiss Cheese' chiral models},'' {\em JHEP} {\bf 0907} (2009) 074,
\href{http://www.arXiv.org/abs/0811.4599}{{\tt 0811.4599}}.

\bibitem{Bobkov:2010rf}
K.~Bobkov, V.~Braun, P.~Kumar, and S.~Raby, ``{Stabilizing All Kahler Moduli in
  Type IIB Orientifolds},'' {\em JHEP} {\bf 1012} (2010) 056,
\href{http://www.arXiv.org/abs/1003.1982}{{\tt 1003.1982}}.

\bibitem{Grimm:2011dj}
T.~W. Grimm, M.~Kerstan, E.~Palti, and T.~Weigand, ``{On Fluxed Instantons and
  Moduli Stabilisation in IIB Orientifolds and F-theory},'' {\em Phys.Rev.}
  {\bf D84} (2011) 066001,
\href{http://www.arXiv.org/abs/1105.3193}{{\tt 1105.3193}}.

\bibitem{Cicoli:2011qg}
M.~Cicoli, C.~Mayrhofer, and R.~Valandro, ``{Moduli Stabilisation for Chiral
  Global Models},'' {\em JHEP} {\bf 1202} (2012) 062,
\href{http://www.arXiv.org/abs/1110.3333}{{\tt 1110.3333}}.

\bibitem{Balasubramanian:2012wd}
V.~Balasubramanian, P.~Berglund, V.~Braun, and I.~Garcia-Etxebarria, ``{Global
  embeddings for branes at toric singularities},'' {\em JHEP} {\bf 1210} (2012)
  132,
\href{http://www.arXiv.org/abs/1201.5379}{{\tt 1201.5379}}.

\bibitem{Cicoli:2012vw}
M.~Cicoli, S.~Krippendorf, C.~Mayrhofer, F.~Quevedo, and R.~Valandro,
  ``{D-Branes at del Pezzo Singularities: Global Embedding and Moduli
  Stabilisation},'' {\em JHEP} {\bf 1209} (2012) 019,
\href{http://www.arXiv.org/abs/1206.5237}{{\tt 1206.5237}}.

\bibitem{Cicoli:2013mpa}
M.~Cicoli, S.~Krippendorf, C.~Mayrhofer, F.~Quevedo, and R.~Valandro, ``{D3/D7
  Branes at Singularities: Constraints from Global Embedding and Moduli
  Stabilisation},''
\href{http://www.arXiv.org/abs/1304.0022}{{\tt 1304.0022}}.

\bibitem{Gao:2013rra}
X.~Gao and P.~Shukla, ``{Stabilization of Odd Axions in LARGE Volume
  Scenario},''
\href{http://www.arXiv.org/abs/1307.1141}{{\tt 1307.1141}}.

\bibitem{2000math......8011D}
R.~{Donagi}, B.~{Ovrut}, T.~{Pantev}, and D.~{Waldram}, ``{Spectral involutions
  on rational elliptic surfaces},'' {\em ArXiv Mathematics e-prints} (Aug.,
  2000) \href{http://www.arXiv.org/abs/arXiv:math/0008011}{{\tt
  arXiv:math/0008011}}.

\bibitem{Ovrut:2003zj}
B.~A. Ovrut, T.~Pantev, and R.~Reinbacher, ``{Invariant homology on standard
  model manifolds},'' {\em JHEP} {\bf 0401} (2004) 059,
\href{http://www.arXiv.org/abs/hep-th/0303020}{{\tt hep-th/0303020}}.

\bibitem{Blumenhagen:2008zz}
R.~Blumenhagen, V.~Braun, T.~W. Grimm, and T.~Weigand, ``{GUTs in Type IIB
  Orientifold Compactifications},'' {\em Nucl.Phys.} {\bf B815} (2009) 1--94,
\href{http://www.arXiv.org/abs/0811.2936}{{\tt 0811.2936}}.

\bibitem{Cicoli:2011it}
M.~Cicoli, M.~Kreuzer, and C.~Mayrhofer, ``{Toric K3-Fibred Calabi-Yau
  Manifolds with del Pezzo Divisors for String Compactifications},'' {\em JHEP}
  {\bf 1202} (2012) 002,
\href{http://www.arXiv.org/abs/1107.0383}{{\tt 1107.0383}}.

\bibitem{Collinucci:2009uh}
A.~Collinucci, ``{New F-theory lifts. II. Permutation orientifolds and enhanced
  singularities},'' {\em JHEP} {\bf 1004} (2010) 076,
\href{http://www.arXiv.org/abs/0906.0003}{{\tt 0906.0003}}.

\bibitem{Kreuzer:2000xy}
M.~Kreuzer and H.~Skarke, ``{Complete classification of reflexive polyhedra in
  four-dimensions},'' {\em Adv.Theor.Math.Phys.} {\bf 4} (2002) 1209--1230,
\href{http://www.arXiv.org/abs/hep-th/0002240}{{\tt hep-th/0002240}}.

\bibitem{Gray:2012jy}
J.~Gray, Y.-H. He, V.~Jejjala, B.~Jurke, B.~D. Nelson, {\em et al.},
  ``{Calabi-Yau Manifolds with Large Volume Vacua},'' {\em Phys.Rev.} {\bf D86}
  (2012) 101901,
\href{http://www.arXiv.org/abs/1207.5801}{{\tt 1207.5801}}.

\bibitem{Blumenhagen:2008ji}
R.~Blumenhagen and M.~Schmidt-Sommerfeld, ``{Power Towers of String Instantons
  for N=1 Vacua},'' {\em JHEP} {\bf 0807} (2008) 027,
\href{http://www.arXiv.org/abs/0803.1562}{{\tt 0803.1562}}.

\bibitem{Petersson:2010qu}
C.~Petersson, P.~Soler, and A.~M. Uranga, ``{D-instanton and polyinstanton
  effects from type I' D0-brane loops},'' {\em JHEP} {\bf 1006} (2010) 089,
\href{http://www.arXiv.org/abs/1001.3390}{{\tt 1001.3390}}.

\bibitem{Cicoli:2011ct}
M.~Cicoli, F.~G. Pedro, and G.~Tasinato, ``{Poly-instanton Inflation},'' {\em
  JCAP} {\bf 1112} (2011) 022,
\href{http://www.arXiv.org/abs/1110.6182}{{\tt 1110.6182}}.

\bibitem{Blumenhagen:2012Poly1}
R.~Blumenhagen, X.~Gao, T.~Rahn, and P.~Shukla, ``{A Note on Poly-Instanton
  Effects in Type IIB Orientifolds on Calabi-Yau Threefolds},'' {\em JHEP} {\bf
  1206} (2012) 162,
\href{http://www.arXiv.org/abs/1205.2485}{{\tt 1205.2485}}.

\bibitem{Blumenhagen:2012ue}
R.~Blumenhagen, X.~Gao, T.~Rahn, and P.~Shukla, ``{Moduli Stabilization and
  Inflationary Cosmology with Poly-Instantons in Type IIB Orientifolds},''
\href{http://www.arXiv.org/abs/1208.1160}{{\tt 1208.1160}}.

\bibitem{Gao:2013hn}
X.~Gao and P.~Shukla, ``{On Non-Gaussianities in Two-Field Poly-Instanton
  Inflation},'' {\em JHEP} {\bf 1303} (2013) 061,
\href{http://www.arXiv.org/abs/1301.6076}{{\tt 1301.6076}}.

\bibitem{Bianchi:2011qh}
M.~Bianchi, A.~Collinucci, and L.~Martucci, ``{Magnetized E3-brane instantons
  in F-theory},'' {\em JHEP} {\bf 1112} (2011) 045,
\href{http://www.arXiv.org/abs/1107.3732}{{\tt 1107.3732}}.

\bibitem{Louis:2012nb}
J.~Louis, M.~Rummel, R.~Valandro, and A.~Westphal, ``{Building an explicit de
  Sitter},'' {\em JHEP} {\bf 1210} (2012) 163,
\href{http://www.arXiv.org/abs/1208.3208}{{\tt 1208.3208}}.

\bibitem{Cicoli:2011yy}
M.~Cicoli, C.~Burgess, and F.~Quevedo, ``{Anisotropic Modulus Stabilisation:
  Strings at LHC Scales with Micron-sized Extra Dimensions},'' {\em JHEP} {\bf
  1110} (2011) 119,
\href{http://www.arXiv.org/abs/1105.2107}{{\tt 1105.2107}}.

\bibitem{Lust:2013kt}
D.~Lust and X.~Zhang, ``{Four Kahler Moduli Stabilisation in type IIB
  Orientifolds with K3-fibred Calabi-Yau threefold compactification},''
\href{http://www.arXiv.org/abs/1301.7280}{{\tt 1301.7280}}.

\bibitem{Gao:2013}
X.~Gao, ``{Scanning tools for toric analyze},''.

\bibitem{Blumenhagen:2010pv}
R.~Blumenhagen, B.~Jurke, T.~Rahn, and H.~Roschy, ``{Cohomology of Line
  Bundles: A Computational Algorithm},'' {\em J. Math. Phys.} {\bf 51} (2010)
  103525, \href{http://www.arXiv.org/abs/1003.5217}{{\tt 1003.5217}}.

\bibitem{cohomCalg:Implementation}
``{cohomCalg package}.'' Download link:
  http://wwwth.mppmu.mpg.de/members/blumenha/cohomcalg/, 2010.
\newblock High-performance line bundle cohomology computation based on
  \cite{Blumenhagen:2010pv}.

\bibitem{Kreuzer:2002uu}
M.~Kreuzer and H.~Skarke, ``{PALP: A Package for analyzing lattice polytopes
  with applications to toric geometry},'' {\em Comput.Phys.Commun.} {\bf 157}
  (2004) 87--106,
\href{http://www.arXiv.org/abs/math/0204356}{{\tt math/0204356}}.

\bibitem{Braun:2012vh}
A.~P. Braun, J.~Knapp, E.~Scheidegger, H.~Skarke, and N.-O. Walliser, ``{PALP -
  a User Manual},''
\href{http://www.arXiv.org/abs/1205.4147}{{\tt 1205.4147}}.

\bibitem{DGPS}
G.-M. P. G. S.~H. Decker, W.;~Greuel, ``{\sc Singular} {3-1-6} --- {A} computer
  algebra system for polynomial computations,''. http://www.singular.uni-kl.de.

\bibitem{sage}
W.~Stein {\em et al.}, {\em {S}age {M}athematics {S}oftware ({V}ersion x.y.z)}.
\newblock The Sage Development Team, YYYY.
\newblock {\tt http://www.sagemath.org}.

\bibitem{Hosono:1993qy}
S.~Hosono, A.~Klemm, S.~Theisen, and S.-T. Yau, ``{Mirror symmetry, mirror map
  and applications to Calabi-Yau hypersurfaces},'' {\em Commun.Math.Phys.} {\bf
  167} (1995) 301--350,
\href{http://www.arXiv.org/abs/hep-th/9308122}{{\tt hep-th/9308122}}.

\bibitem{Gao:2012}
X.~Gao, ``{Representation of vector-bundle valued cohomology on
  hypersurface}(Mathematica File),''.

\bibitem{Candelas:1987kf}
P.~Candelas, A.~Dale, C.~Lutken, and R.~Schimmrigk, ``{Complete Intersection
  Calabi-Yau Manifolds},'' {\em Nucl.Phys.} {\bf B298} (1988)
493.

\bibitem{Gray:2013mja}
J.~Gray, A.~S. Haupt, and A.~Lukas, ``{All Complete Intersection Calabi-Yau
  Four-Folds},'' {\em JHEP} {\bf 1307} (2013) 070,
\href{http://www.arXiv.org/abs/1303.1832}{{\tt 1303.1832}}.

\end{thebibliography}\endgroup
\bibliographystyle{utphys}

\end{document}